\documentclass[%
twocolumn,
superscriptaddress,
groupedaddress,
 amsmath,amssymb,
 aps,
prb,
]{revtex4}

\usepackage{hyperref}
\usepackage{graphicx}
\usepackage{amsmath}
\usepackage{sidecap}
\usepackage{amsfonts}
\begin{document}
\title{Residual spin susceptibility in the spin-triplet, orbital-singlet model}
\author{Yue Yu}
\author{Alfred K. C. Cheung}
\affiliation{Department of Physics, Stanford University, Stanford, CA 94305}
\author{S. Raghu}
\affiliation{Department of Physics, Stanford University, Stanford, CA 94305}
\affiliation{SLAC National Accelerator Laboratory, Menlo Park, CA 94025}
\author{D. F. Agterberg}
\affiliation{Department of Physics, University of Wisconsin, Milwaukee, WI 53201}

\begin{abstract}
Nuclear magnetic resonance (NMR) and Knight shift measurements are critical tools in the identification of spin-triplet superconductors. We discuss the effects of spin orbit coupling on the Knight shift and susceptibilities for a variety of spin triplet multi-orbital gap functions with orbital-singlet character and compare their responses to "traditional" single band spin-triplet ($p_x+ip_y$) superconductors. We observe a non-negligible residual spin-susceptibility at low temperature.
\end{abstract}

\maketitle
\section{Introduction}

A working definition for an unconventional superconductor is one whose gap function averaged over the Fermi surface is less than the maximum value of the absolute value of the gap at any point on the Fermi surface. This allows for gaps which do not exhibit isotropic $s$-wave pairing. In particular, the possibility for pairing in odd parity channels allows for \textit{spin-triplet pairing}, the most notable purported example being Sr$_2$RuO$_4$ in which the simplest descriptions involve models with only intra-band pairing. For Sr$_2$RuO$_4$, there is still no consensus on the actual form of the superconducting gap with various experiments showing conflicting results~\cite{scaffidi2017thesis,ishida1998spin,xia1,hicks1,steppke2017}. In light of this, it is useful to consider what other systems might be candidates for realizing spin-triplet pairing. One recent work proposes that a different kind of spin-triplet pairing may be realized in iron-based superconductors that only possess hole pockets~\cite{vafek1}. Based on angle-resolved photoemission spectroscopy and heat capacity measurements, the authors argue in favor of $s$-wave gaps in these materials and present a new mechanism for its realization.

The key ingredients to their proposal are (1) the presence of spin orbit coupling (SOC), and (2) the multi-orbital nature of these systems which introduces the possibility for inter-band paired gap functions~\cite{ramires2,hoshino1}. This results in the stabilization of an even parity orbital-singlet, spin-triplet pairing state. Indeed, the $s$-wave iron-based superconductors are expected to exhibit sizable SOC~\cite{borisenko2016direct} and seem to have the ingredients necessary for the model proposed~\cite{nisson2016nuclear,haverkort1,vafek1,ramires1}. The same pairing state has also been the focus of study using dynamical mean field theory~\cite{klejnberg1999,spalek2001,han2004,hoshino1}. These studies all reveal a pairing instability within the strong coupling limit. 

In order to evaluate the validity of the proposed model for the relevant iron-based supercondcutors, it is crucial to identify the experimental signatures in nuclear magnetic resonance (NMR) and Knight shift measurements -- key experimental testing grounds for unconventional superconductors~\cite{ishida1998spin, mackenzie2003superconductivity,abragam1961principles,curro2009nuclear,rigamonti1998basic}. To this end, in this article we compare the results in Knight shift between the well-studied single band spin-triplet state $\vec{d}(\vec{k})=(0,0,k_x+ik_y)$ and the inter-band model of Ref.~\onlinecite{vafek1}. In the absence of SOC, we find an invariant Knight shift which stays constant into the superconducting phase in every direction for the inter-band model. This distinguishes it from the intra-band $(k_x+ik_y)\hat{z}$ state, where there is a drop in the Knight shift for a field applied along the $z$ direction \cite{PhysRevLett.92.097001,nisson2016nuclear}. When including SOC, we observe a substantial decrease in spin susceptibilities, which agrees with previous theoretical predictions \cite{vafek1}. However, we observe a non-zero residual spin susceptibility at low temperature, in contrast with zero residual spin susceptibility for intra-band spin-singlet pairing. Our results on spin susceptibility and Knight shift reveal that the pairing state driven by SOC has both intra-band spin-singlet and inter-band spin-triplet properties. 

This article is organized as follows. In Section~\ref{sect:model}, we introduce our mean field theory model involving pairing in the orbital singlet, spin triplet channel. In Section~\ref{sect:formalism}, we explain how observables in NMR and Knight shift experiments can be calculated within our model. Our results are discussed in Section~\ref{result}. Concluding remarks appear in Section~\ref{sect:conc}.

\section{The model} \label{sect:model}
We consider a three band model for $d$-orbital electrons with tetragonal symmetry. The Hamiltonian is given by $H=H_0+H_{SOC}+H_{BCS}$, where:
\begin{equation}
H_0=\sum_{\vec{k},a,b,\sigma}h_0^{ab}(\vec{k})c_{\vec{k}a\sigma}^{\dagger}c_{\vec{k}b\sigma},
\end{equation}
and:
\begin{equation}
h_0^{ab}(\vec{k})=\left[\begin{array}{ccc}\epsilon_{yz}(\vec{k})& V(\vec{k}) & 0 \\ V(\vec{k})&\epsilon_{xz}(\vec{k})& 0 \\ 0 & 0 &\epsilon_{xy}(\vec{k})\end{array}\right],
\end{equation}
is the kinetic energy part of the tight binding model with hybridization between $xz$ and $yz$ orbitals. $c_{\vec{k}a\sigma}^\dag \left(c_{\vec{k}a\sigma} \right)$ is the creation(annihilation) operator of electrons in orbital $a=yz$, $xz$, or $yz$ and spin $\sigma = \uparrow, \downarrow$. We consider a quasi-two dimensional material, where the dispersion in the $z$-direction is neglected. Here, the form and value of the unhybridized dispersions and the hybridization potential are given in Ref.~\onlinecite{band}:
\begin{equation}
\begin{split}
&\epsilon_{yz}(\vec{k})=-\epsilon'-2t\cos{k_y}-2t_\perp\cos{k_x}\\
&\epsilon_{xz}(\vec{k})=-\epsilon'-2t\cos{k_x}-2t_\perp\cos{k_y}\\
&\epsilon_{xy}(\vec{k})=-\epsilon-2t'(\cos{k_x}+\cos{k_y})+4t''\cos{k_x}\cos{k_y}\\
&V(\vec{k})=-2V\sin{k_x}\sin{k_y},
\end{split}
\end{equation}
which was originally proposed for $\text{Sr}_2\text{RuO}_4$. However, we use this model only as a specific example; our analysis is not  limited to this particular material. We choose our unit of energy to be $t$ in the following analysis. Without spin orbit coupling (SOC) and superconductivity, the band structure from the diagonalization of $H_0$ is given in Ref.~\onlinecite{band}. The minimal band gap at the Fermi surface $\Delta_{\textrm{band}}\approx0.05t$ is between the $d_{xy}$ band and one of the hybridized bands. We add the spin orbit coupling $H_{SOC}=\lambda\vec{L}\cdot\vec{S}$. The form of the BCS interaction we use is: 
\begin{widetext}
\begin{equation}
\begin{split}
H_{\text{BCS}}=-\sum_{\vec{k},\vec{k}',a,b,\lbrace{\sigma_i}\rbrace}V_{ab\sigma_1\sigma_2\sigma_3\sigma_4}(\vec{k},\vec{k}'){c}_{-\vec{k}a\sigma_1}{c}_{\vec{k}b\sigma_2}{c}_{\vec{k}' b\sigma_3}^{\dagger}c_{-\vec{k}' a\sigma_4}^{\dagger},
\end{split}
\end{equation}
\end{widetext}
Using a mean field decomposition, we can calculate the gap function:
\begin{equation}
\begin{split}
\Delta_{ab\sigma_3\sigma_4}(\vec{k'})=\sum_{\vec{k}\sigma_1\sigma_2}V_{ab\sigma_1\sigma_2\sigma_3\sigma_4}(\vec{k},\vec{k'})\langle{c}_{-\vec{k}a\sigma_1}{c}_{\vec{k}b\sigma_2}\rangle.
\end{split}
\end{equation}
Thus we obtain the mean field Hamiltonian, which is now rewritten into Bogoliubov-de-Gennes (BdG) form:  
\begin{align}
&\Psi_{\vec{k}}^{\dagger}\equiv[c_{\vec{k}X\uparrow}^{\dagger},c_{\vec{k}X\downarrow}^{\dagger},c_{\vec{k}Y\uparrow}^{\dagger},c_{\vec{k}Y\downarrow}^{\dagger},c_{\vec{k}Z\uparrow}^{\dagger},c_{\vec{k}Z\downarrow}^{\dagger}]\nonumber\\
&H_{MF}=\sum_{\vec{k}|k_x>{0}}[\Psi_{\vec{k}}^{\dagger},\Psi_{\vec{-k}}^{T}]h_{BdG}(\vec{k})\left[\begin{array}{c} \Psi_{\vec{k}}\\\Psi_{\vec{-k}}^* \end{array}\right]\nonumber\\
&h_{BdG}(\vec{k})=\left(\begin{array}{cc}\hat{h}(\vec{k})&\hat{\Delta}(\vec{k})\\{\hat{\Delta}(\vec{k})^\dagger}&-\hat{h}^T(\vec{-k})\end{array}\right);\nonumber\\
&\hat{h}(\vec{k})=h_0^{ab}(\vec{k})\otimes\sigma_0\nonumber\\
&+\lambda(L_x\otimes\sigma_x+L_y\otimes\sigma_y+L_z\otimes\sigma_z),
\end{align}
as a 12 by 12 matrix. Here $\sigma_{x,y,z,0}$ are the Pauli matrices and identity matrix in spin space. For simplicity, the orbital angular momentum operators $(L_{a})^{bc}=-i\epsilon_{abc}$ are assumed to be the same as for electrons with $L=1$.

For the case of intra-band pairing, the superconducting order parameter can be rewritten as $\Delta(\vec{k})=(I_3)\otimes{i\sigma_y}{\vec{d}(\vec{k})\cdot\vec{\sigma}}$, where $\vec{\sigma}=(\sigma_x,\sigma_y,\sigma_z)$ and $I_3$ is the 3 by 3 identity matrix in orbital space\cite{PhysRevLett.92.097001}. Additionally, the multi-band nature of the model introduces the possibility for inter-band coupled gap functions. We consider a BCS gap $\Delta_{BCS}=0.01t$, which is of the order of the band gap near $E_F$. Thus, the inter-band pairing cannot be fully neglected. In this calculation, we consider a local model ($\vec{k}$-independent model) of orbital-singlet and spin-triplet pairing, which has the following gap function, and the corresponding $V_{ab\sigma_1\sigma_2\sigma_3\sigma_4}(\vec{k},\vec{k}')$:
\begin{align}
&V_{ab\sigma_1\sigma_2\sigma_3\sigma_4}(\vec{k},\vec{k}')=g\, f(\vec{k})^\dagger_{ba\sigma_2\sigma_1}f(\vec{k'})_{ab\sigma_3\sigma_4}\nonumber\\
&\hat{\Delta}_{ab\sigma_1\sigma_2}(\vec{k})=\Delta\,{f}_{ab\sigma_1\sigma_2}(\vec{k})\nonumber\\
&f(\vec{k})\equiv{L_z}\otimes(i\sigma_z\sigma_y),
\label{e7}
\end{align}
coupling $yz$ and $xz$ orbitals. Since the induced intraband spin-singlet state does not depend on a specific direction, the results on susceptibilities do not show qualitative differences in different directions, which will be confirmed in the next section. Thus, other pairings including $(L_x\otimes\sigma_x+L_y\otimes\sigma_y)i\sigma_y$ will give similar results. The gap function is even parity, with time reversal symmetry and inversion symmetry. Note that other pairing states (e.g. intra-band pairings or other forms of inter-band pairing) are not considered here. The two electrons comprising the Cooper pair form an orbital singlet using the $xz$ and $yz$ orbitals. In spin space, they form a triplet pair with total spin rotating within the $x-y$ plane. The overall Cooper pair is odd under particle exchange. The $\vec{d}$-vector of this triplet-spin pairing contains only a $z$-component. If we turn on the SOC, this inter-band pair could develop an intra-band spin-singlet component \cite{vafek1}. The finite intra-band pairing as induced by SOC helps increase the superconducting transition temperature, as we will explain in Sec. {\ref{result}}.

\section{NMR and Knight Shift} \label{sect:formalism}
We now consider observables in NMR experiments and the Knight shift, which are key experimental techniques in identifying spin triplet superconductors \cite{abragam1961principles,rigamonti1998basic}. Understanding the Knight shift experiment is vital in distinguishing between different types of gap functions \cite{abrikosov1962spin,PhysRevLett.3.325,PhysRevLett.92.097001}. We will first summarize the theoretical background of this experiment. Then we will show results for the inter-band paired state under different SOC strengths.

In atomic physics, the field induced non-zero spin and orbit angular momentum of the electrons generate a hyperfine field experienced by the nuclear spin \citep{NMR}:
\begin{widetext}
\begin{equation}
\begin{split}
\vec{B}_{\text{hf}}=-2\mu_0\mu_B\langle{r}^{-3}\rangle(\vec{L}+\xi{L}(L+1)\vec{S}-\frac{3}{2}\xi[\vec{L}(\vec{L}\cdot\vec{S})+(\vec{L}\cdot\vec{S})\vec{L}]),
\label{eq:Bhf}
\end{split}
\end{equation}
\end{widetext}
which leads to the Knight shift in the NMR response. The hyperfine field can be decomposed into orbital and spin contributions. The orbital angular momentum generates a current and hence, a magnetic field, contributing to the first term of the hyperfine field in Eq.~\ref{eq:Bhf}. The dipole-dipole interaction between electron spin and nuclear spin leads to the remaining two terms in the hyperfine field, under the approximation known as the Equivalent Operator Method \cite{chivalue}. Given the atomic wavefunction of an electron with angular momentum $l$, the strength of the dipole-dipole interaction is $\xi=2/[(2l-1)(2l+3)]$\cite{chivalue}. In the following calculations, we choose $\ell = 2$ which gives $\xi=2/21$. The Fermi contact interaction will also contribute to the hyperfine field, which is neglected here for $L\neq{0}$ systems. 

Knight shift tensor $\overleftrightarrow{K}$ is then determined by the hyperfine field through 
\begin{equation}
\begin{split}
\vec{B}_{\text{hf}}=\overleftrightarrow{K}\cdot\vec{B}.
\label{ks}
\end{split}
\end{equation}
Here $\vec{B}$ is the external magnetic field. The spin and orbital contribution towards the diagonal elements of the Knight shift is directly related to the spin and orbital susceptibilities. Under spin orbit coupling, $\vec{L}$ and $\vec{S}$ are no longer good quantum numbers, and we will take the expectation value of the hyperfine field operator in the simulation. The orbital contribution, which is proportional to the orbital magnetic susceptibility, does not change dramatically upon entering the superconducting phase. This is because in the Kubo formula for orbital susceptibility, orbital angular momentum couples states with different energy, so the energy shift by superconductivity will not have a strong effect. The orbital contribution can be extracted from measurements within the normal state and then substracted in the superconducting Knight shift \cite{abragam1961principles,rigamonti1998basic}. The spin contribution has a similar behavior as spin susceptibility and acts as a key feature for distinguishing between spin-singlet and spin-triplet gap functions. In the following simulations, we numerically diagonalize the BdG Hamiltonian and obtain the susceptibilities and the diagonal elements of the Knight shift tensor.

\section{Results and discussion} \label{result}
The gap equation can now be solved numerically, and the parameter $\Delta$ in Eq.~\ref{e7} is obtained self-consistently for a given $g$ and temperature $T$. After obtaining $\Delta$, the susceptibilities and Knight shift are calculated from the Kubo formula. We now compare the Knight shift results for the single band $\vec{d}(\vec{k})=(0,0,k_x+ik_y)$ state and the inter-band orbital-singlet, spin-triplet state.  In order to shed light on the role of SOC in the formation of intra-band pairing, we consider three SOC regimes: $\lambda=0$ (i.e. zero SOC) where only inter-band pairing is present, $\lambda\sim\Delta_{BCS}$ when the SOC strength is comparable to the size of the superconducting gap, and $\lambda\gg\Delta_{BCS}$. The Knight shift and susceptibilities are all normalized to a dimensionless number, for which the Knight shift and susceptibilities in the normal state is unity. 

\begin{figure}[htb]
\centering
\includegraphics[width=4cm]{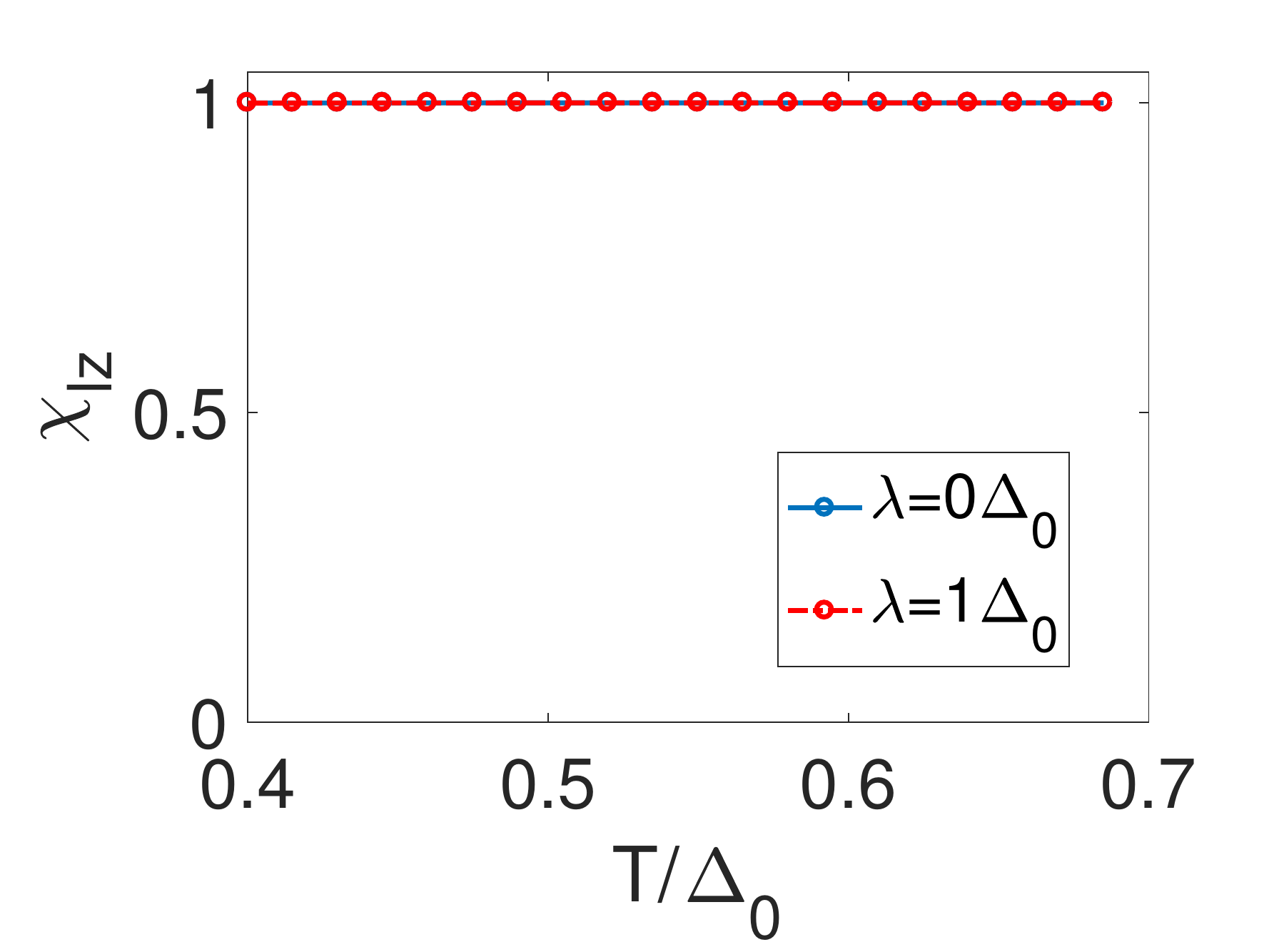}
\includegraphics[width=4cm]{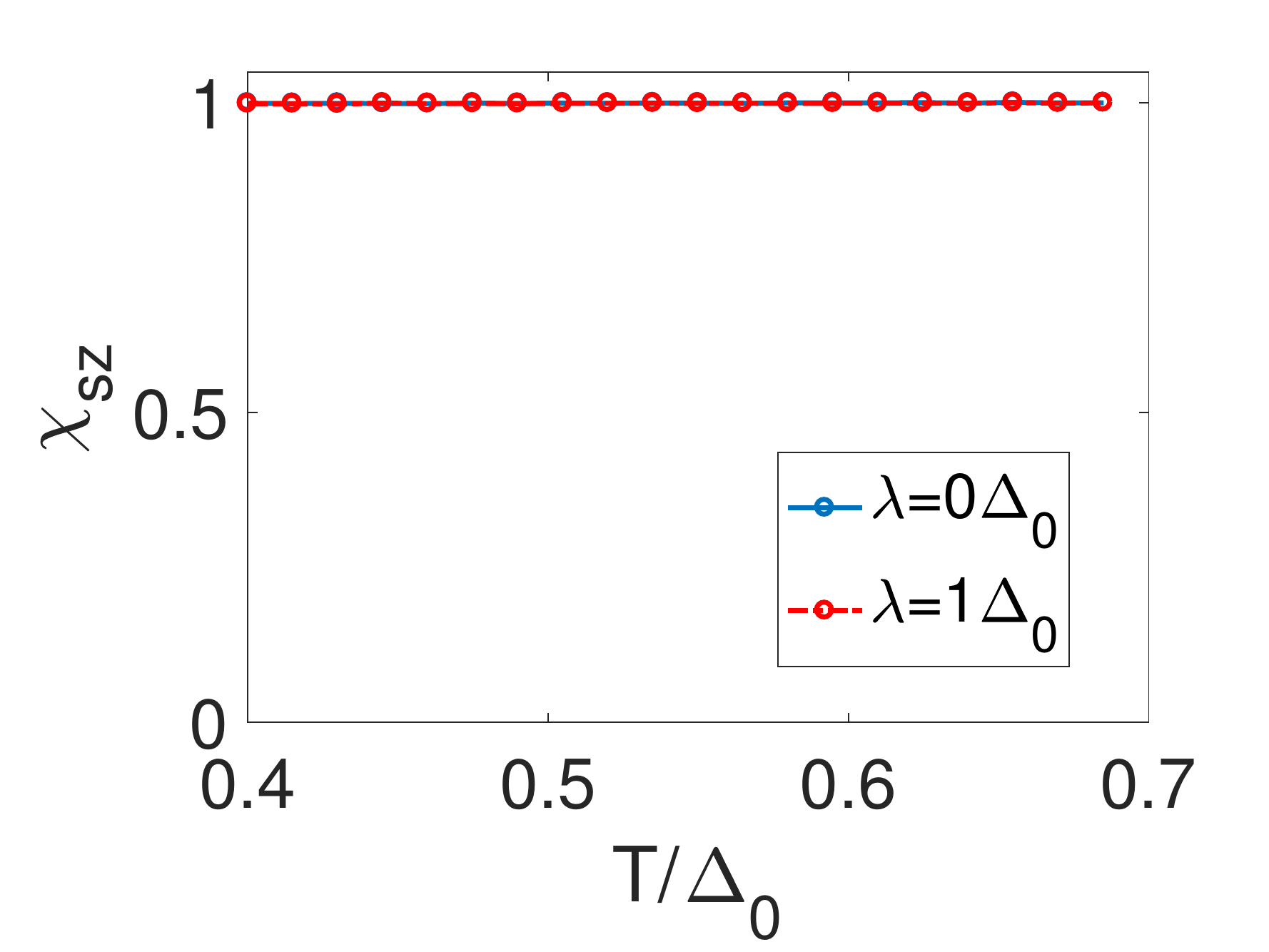}
\includegraphics[width=4cm]{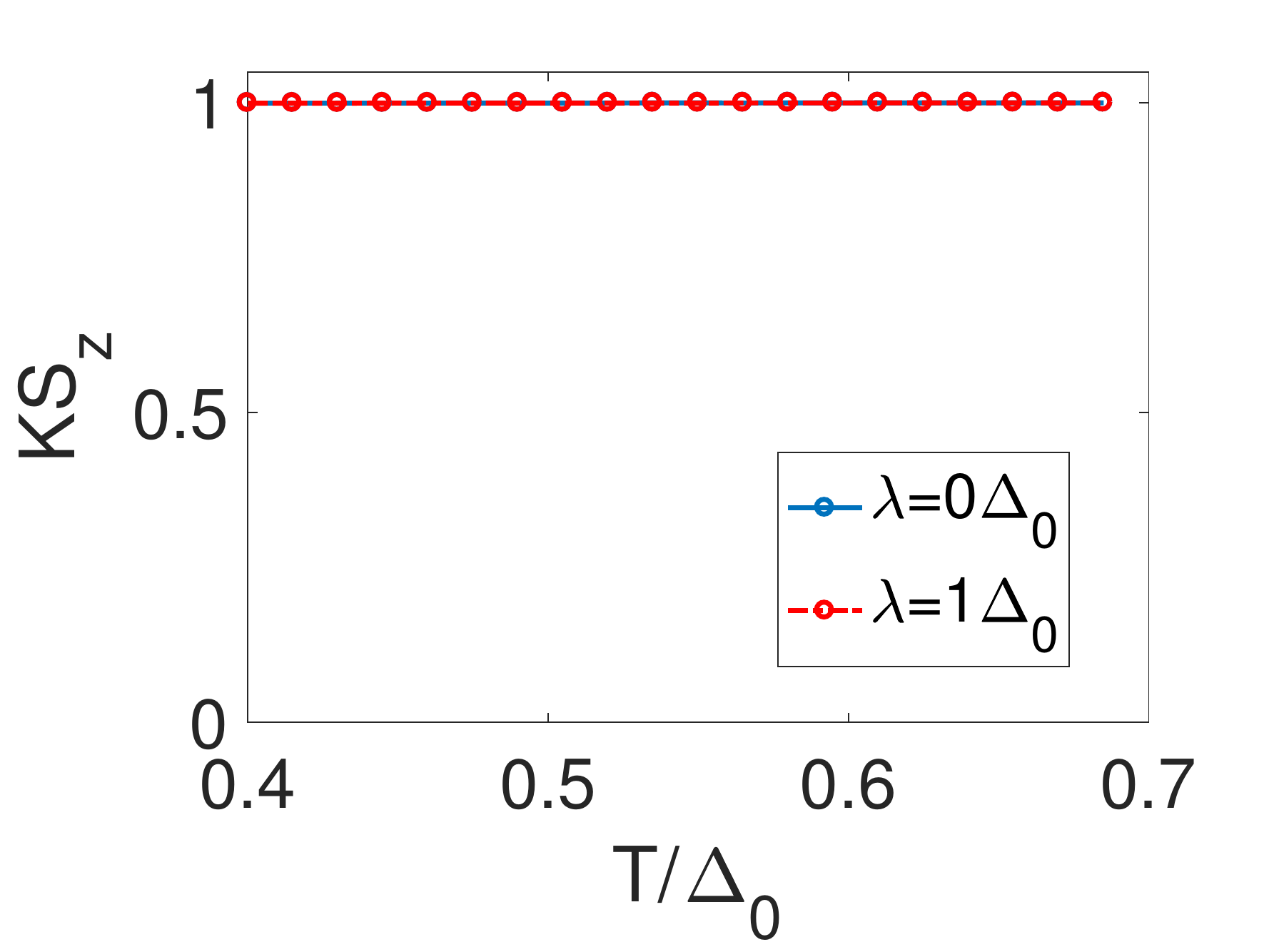}
\includegraphics[width=4cm]{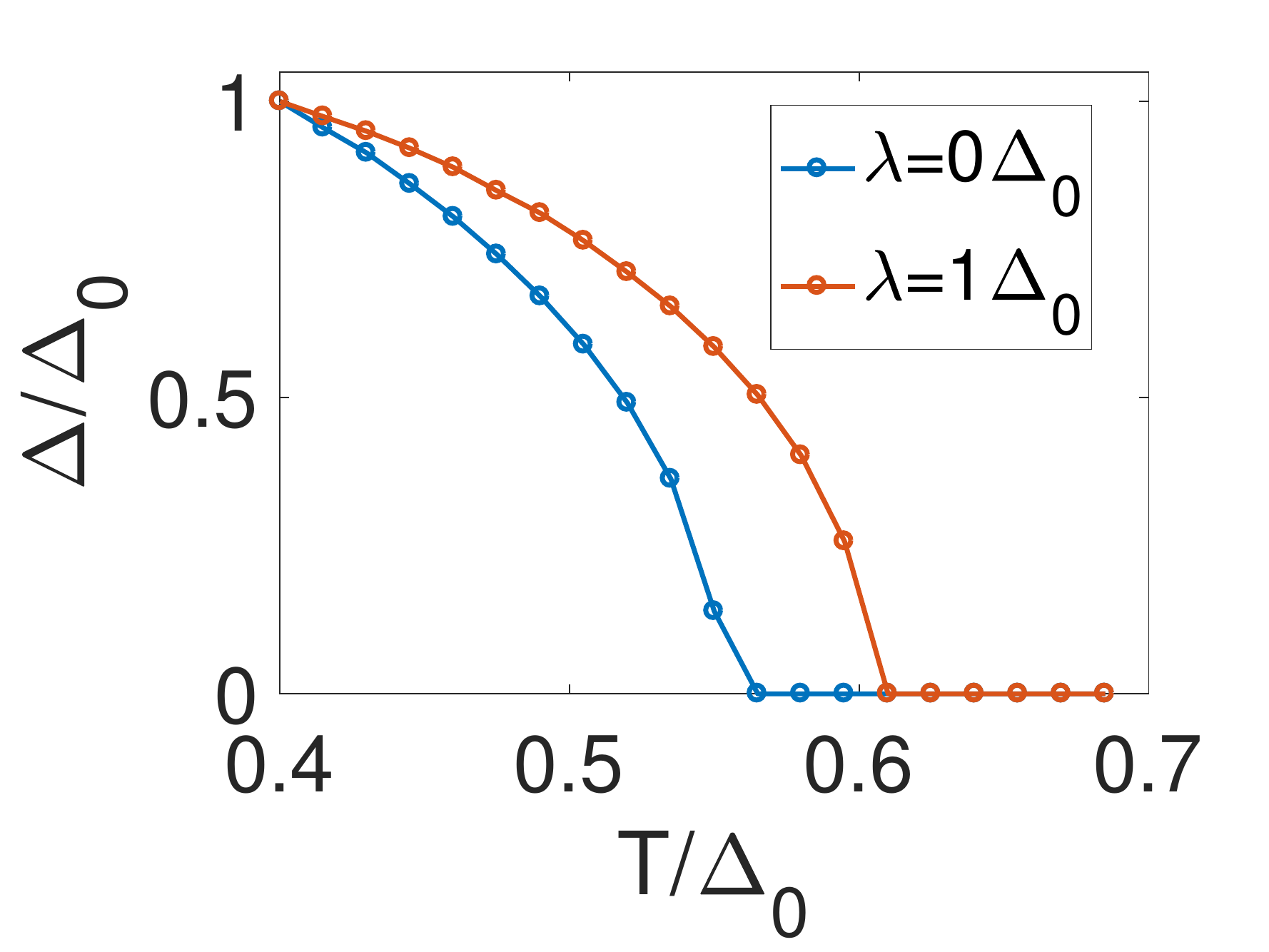}
\caption{Orbital susceptibility, spin susceptibility and gap function for triplet pairing $\Delta(\vec{k})=L_z\otimes{i\sigma_y\sigma_z}$ when applying an out-of-plane magnetic field for $\lambda=0$ and $\lambda=\Delta_{BCS}$. Within the range of the gap function, no drop in susceptibilities and Knight shift is observed. Note that the zero temperature gap function should be larger than $0.01t$, but we only focus on the range where $\Delta_{BCS}\ll\Delta_{band}$.}
\label{F3a} 
\end{figure}

\begin{figure}[htb]
\centering
\includegraphics[width=4cm]{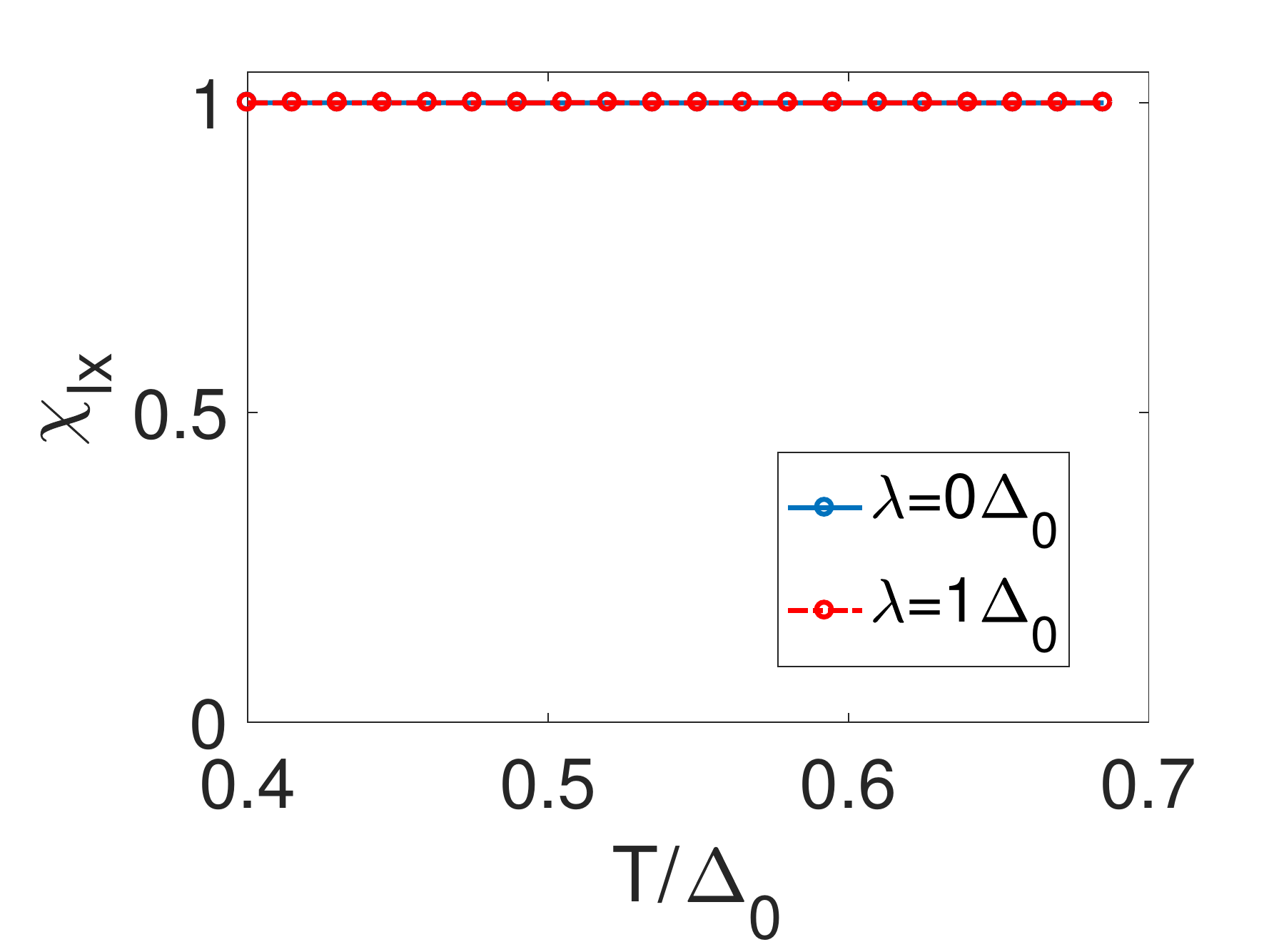}
\includegraphics[width=4cm]{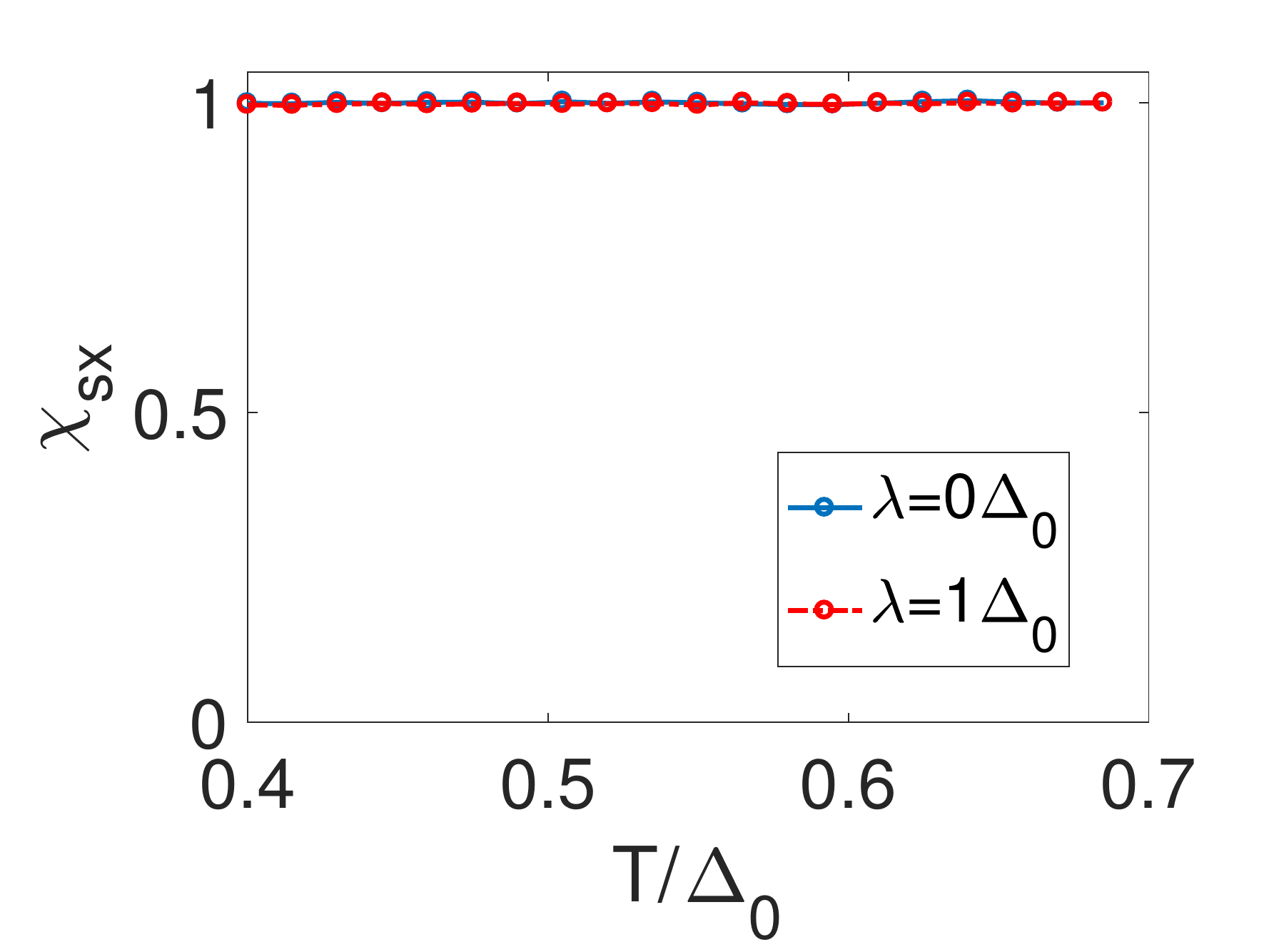}
\includegraphics[width=4cm]{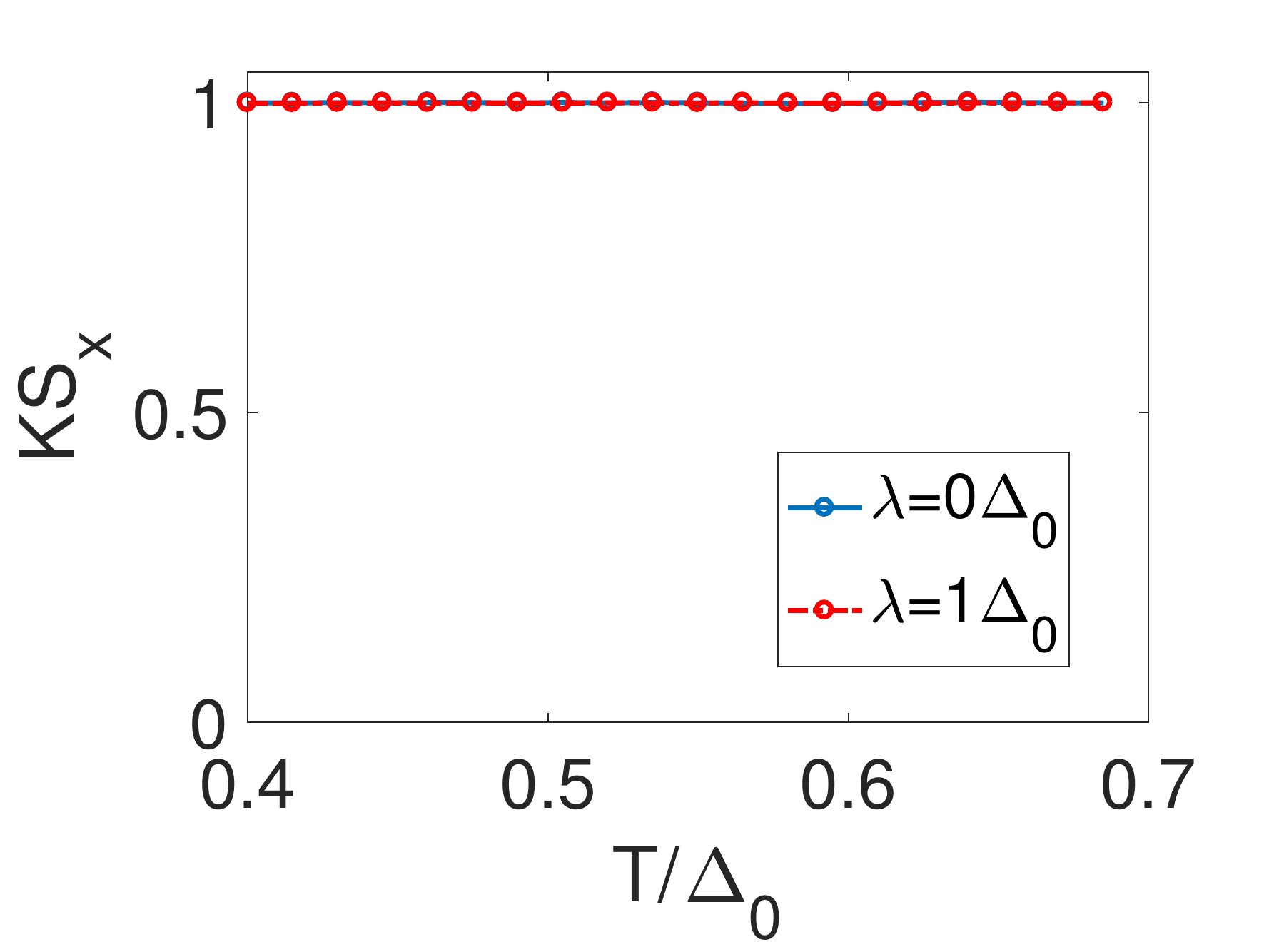}
\includegraphics[width=4cm]{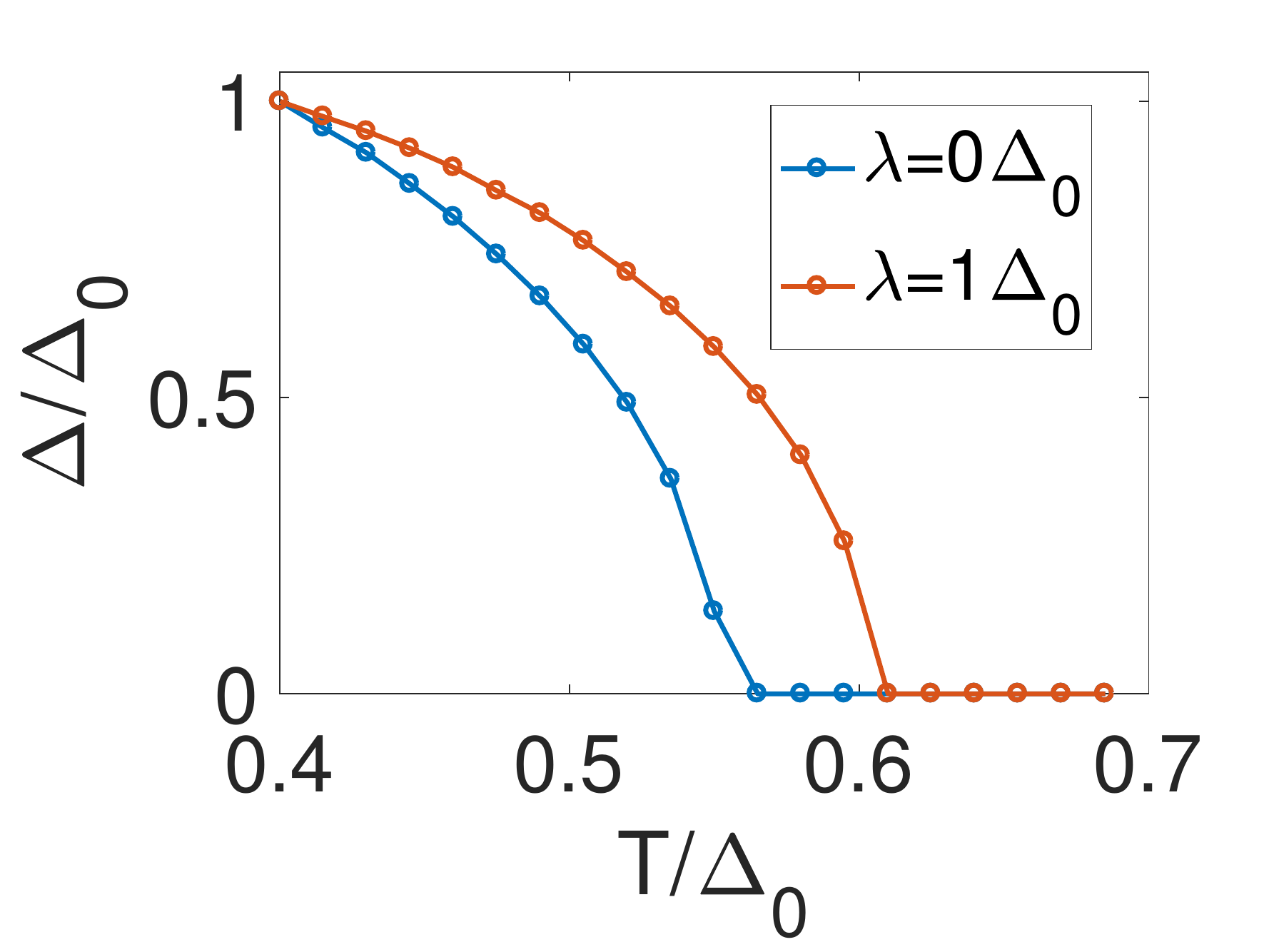}
\caption{Orbital susceptibility, spin susceptibility, and gap function for triplet pairing $\Delta(\vec{k})=L_z\otimes{i\sigma_y\sigma_z}$ when applying a magnetic field in the in-plane $x$-direction. The same parameters are applied as in Fig.~\ref{F3a}.}
\label{F3b}
\end{figure}

If the spin orbit coupling is absent, the critical temperature for superconductivity is found to be exponentially small. We present here an unrealistic scenario with a non-zero order parameter $\Delta_{BCS}$ under zero/small SOC, as shown in Fig.~\ref{F3a} and Fig.~\ref{F3b}. This setup can be achieved by choosing a relatively large coupling coefficient $g$ in Eq.~\ref{e7}, which raises the critical temperature to numerically accessible values. This small SOC regime serves only to illustrate the key results, namely the residual spin-susceptibility. Furthermore, to better illustrate the effect of SOC, we choose different $g$ parameters for the two curves $\lambda=0$ and $\lambda=\Delta_0$, such that the superconductivity gap at the lowest temperature reached is fixed to be $\Delta_0=0.01t$.

For purely inter-band pairing without SOC, we numerically obtain the susceptibility and Knight shift. Fig.~\ref{F3a} shows the responses in the presence of an out-of-plane magnetic field while Fig.~\ref{F3b} is in the presence of an in-plane magnetic field. In striking contrast to the intra-band $k_x+ik_y$ model \cite{PhysRevLett.92.097001,nisson2016nuclear,ishida1998spin}, the spin and orbital susceptibilities and the Knight shift show no decrease in the superconducting phase in any direction. We further check the density of states. By redefining the Fermi surface to be the states with energy near the chemical potential $|E-E_F|<\Delta_{BCS}$, we find that there is no dramatic change in the density of states on the Fermi surface. 
\begin{figure}[htb]
\centering
\includegraphics[width=3cm]{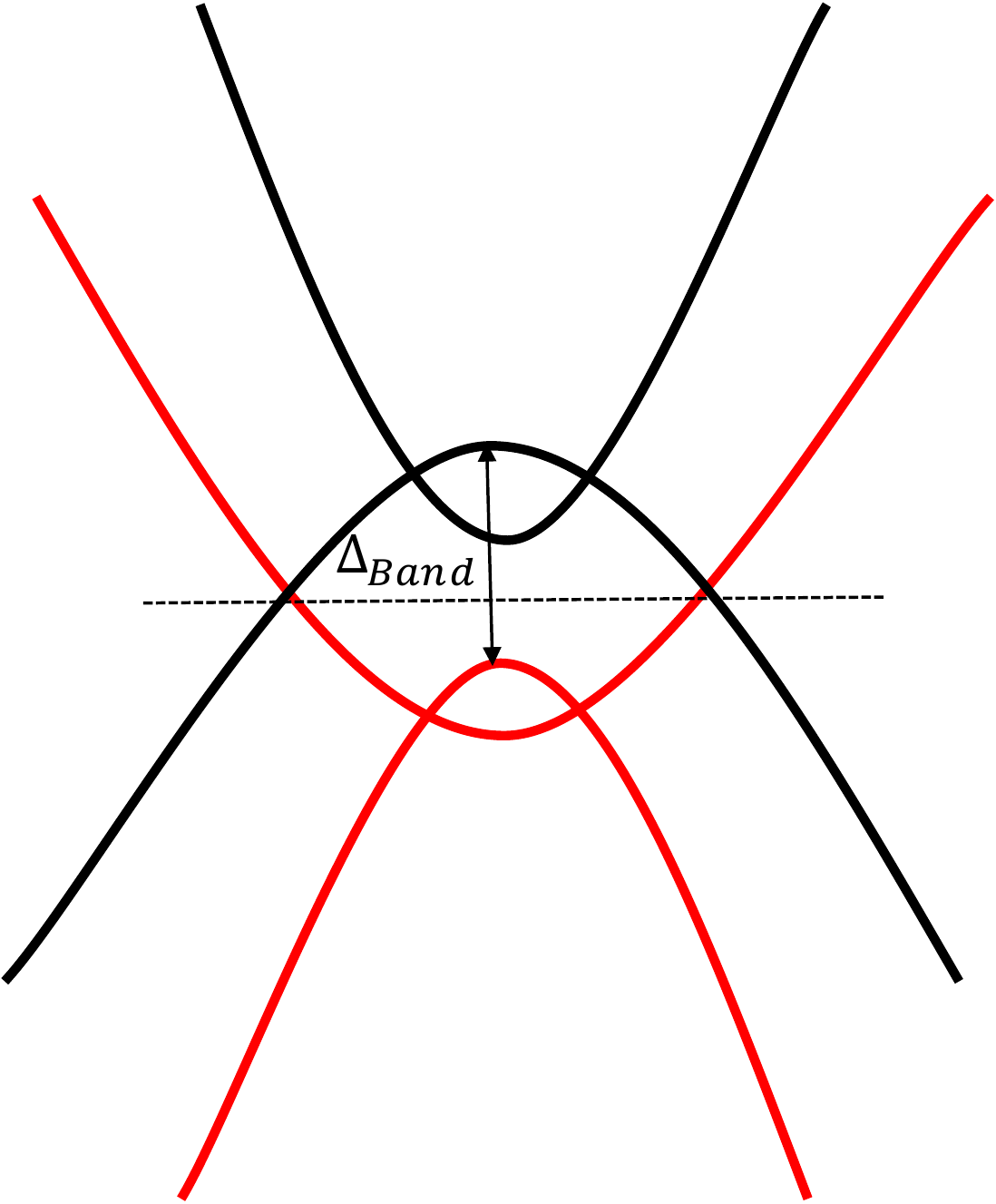}
\includegraphics[width=4cm]{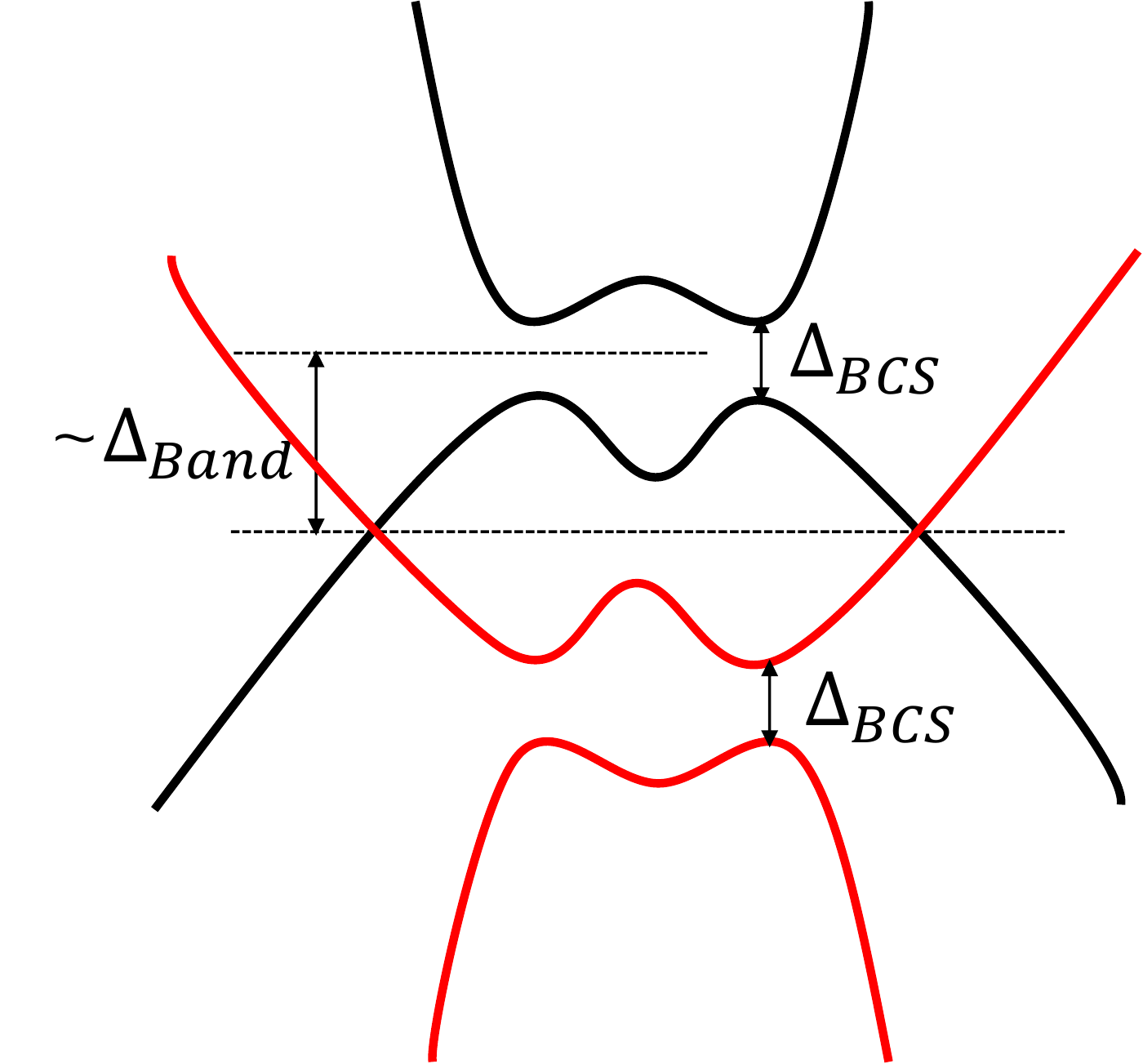}
\caption{Schematic representation of the change of band structure for a two band system with inter-band coupling for (left) no superconductivity and (right) with superconductivity. The distance between the BCS gaps and the Fermi surface is of order $\Delta_{Band}$. If the superconducting order parameter $\Delta_{BCS}$ is much smaller than the energy difference of the two bands $\Delta_{Band}$, then the states near the Fermi surface are approximately unchanged. Thus the density of states at the Fermi surface is unaffected by superconductivity. }
\label{Fermi} 
\end{figure}

Here we provide a qualitative explanation for the above findings. For intra-band pairing, the Cooper pairs are composed of electrons with the same energies, and the contribution to the superconductivity gap is mainly from electrons near the Fermi surface ($|E-E_F|\approx{\Delta_{BCS}}$). When BCS states are formed, a superconducting gap then opens at the Fermi surface. The density of states at the Fermi surface vanishes, and there is a substantial drop in spin susceptibility. In contrast, the inter-band pairing involves electrons at different bands. In a rough approximation, let us assume the pairing only happens at the band crossing (Fig. \ref{Fermi}). When Cooper pairs are formed, a superconducting gap will open above and below the Fermi surface. The distance from the Fermi surface is of order $\Delta_{band}$. If $\Delta_{band}\gg\Delta_{BCS}$, the Fermi surface is then approximately unchanged. Therefore, the inter-band superconductivity does not exhibit any decrease in Knight shift, even though $\vec{d}$ is in $z$-direction. 

As we turn on SOC while still keeping it weak ($\lambda\sim\Delta_{BCS}<\Delta_{Band}$), the SOC is not yet sufficient to generate a considerable intra-band spin-singlet pairing. As a result, the decrease spin-susceptibility and Knight shift remains small (red curves in Fig.~\ref{F3a}and Fig.~\ref{F3b}). However, we observe a higher critical temperature, which is due to a reduction of the minimal band gap by SOC. 

\begin{figure}[htb]
\centering
\includegraphics[width=4cm]{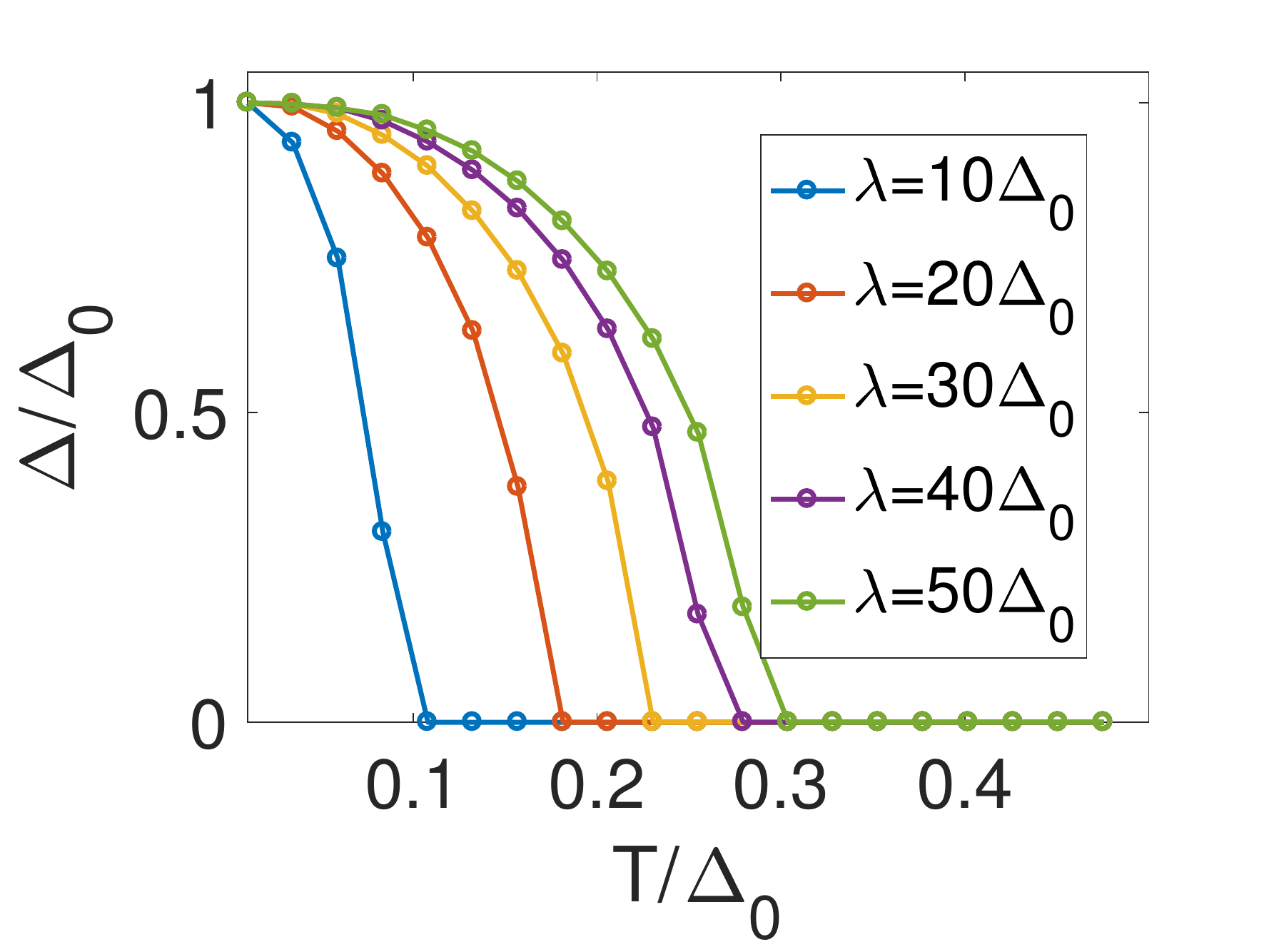}
\includegraphics[width=4cm]{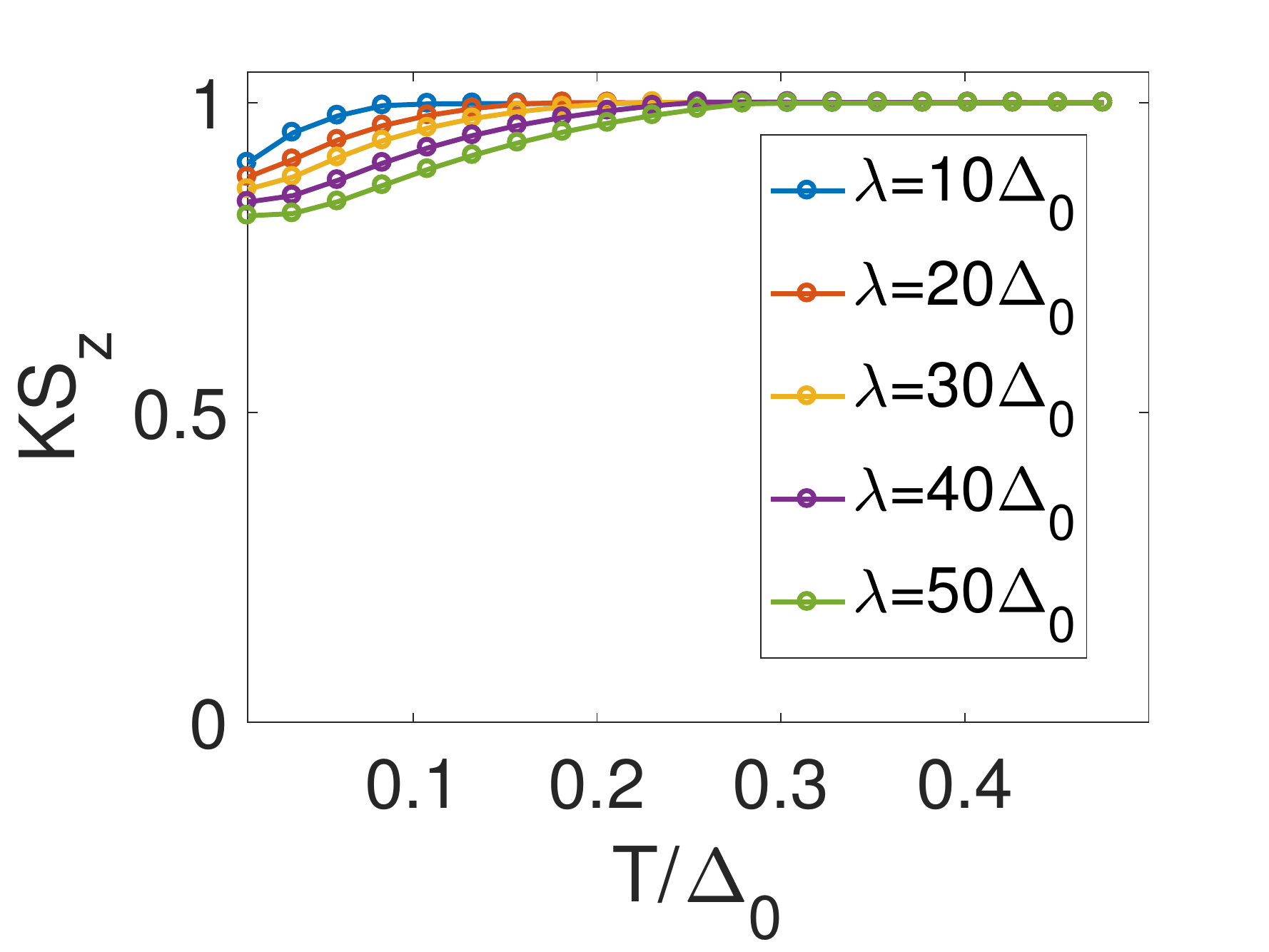}
\includegraphics[width=4cm]{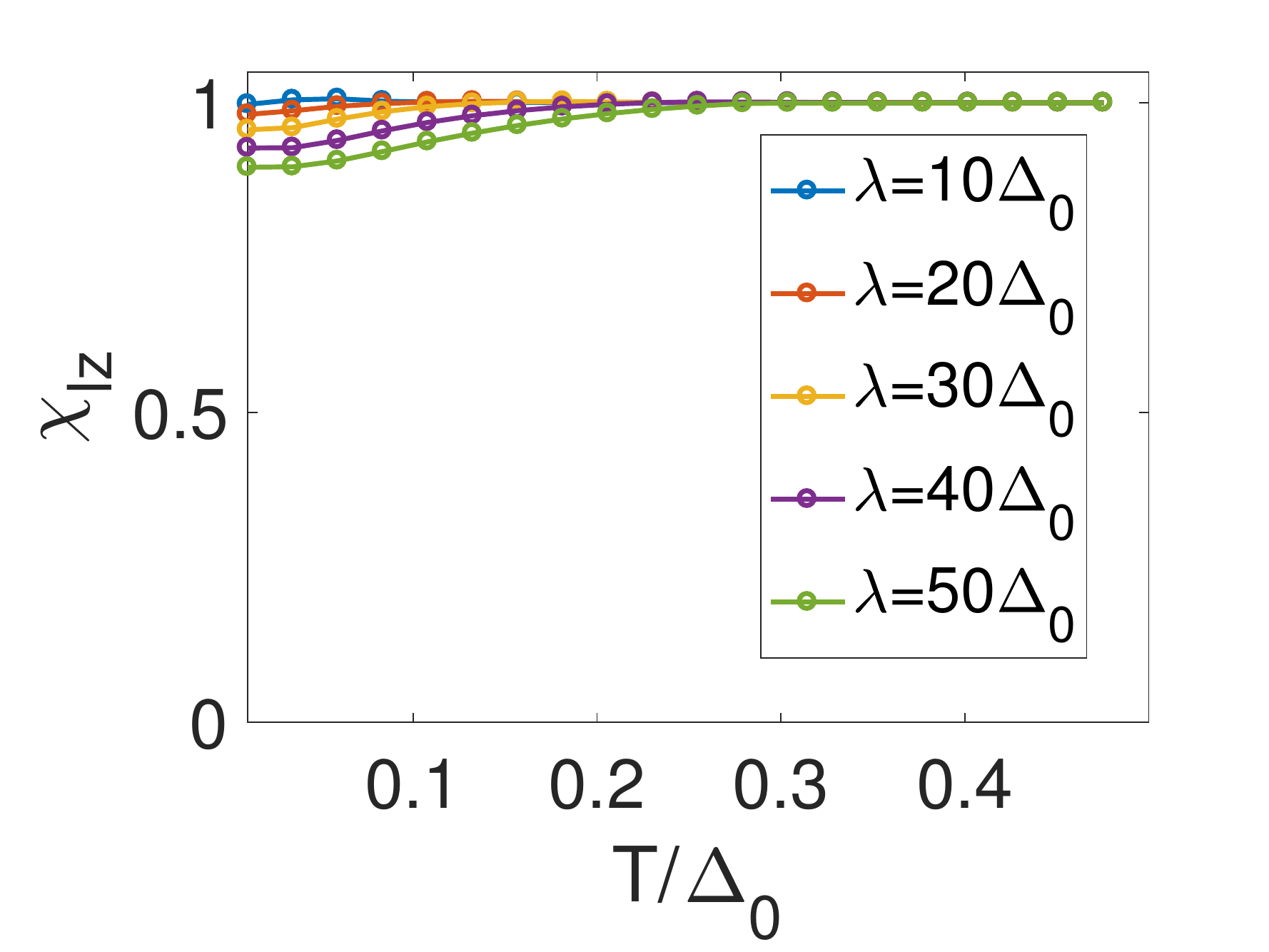}
\includegraphics[width=4cm]{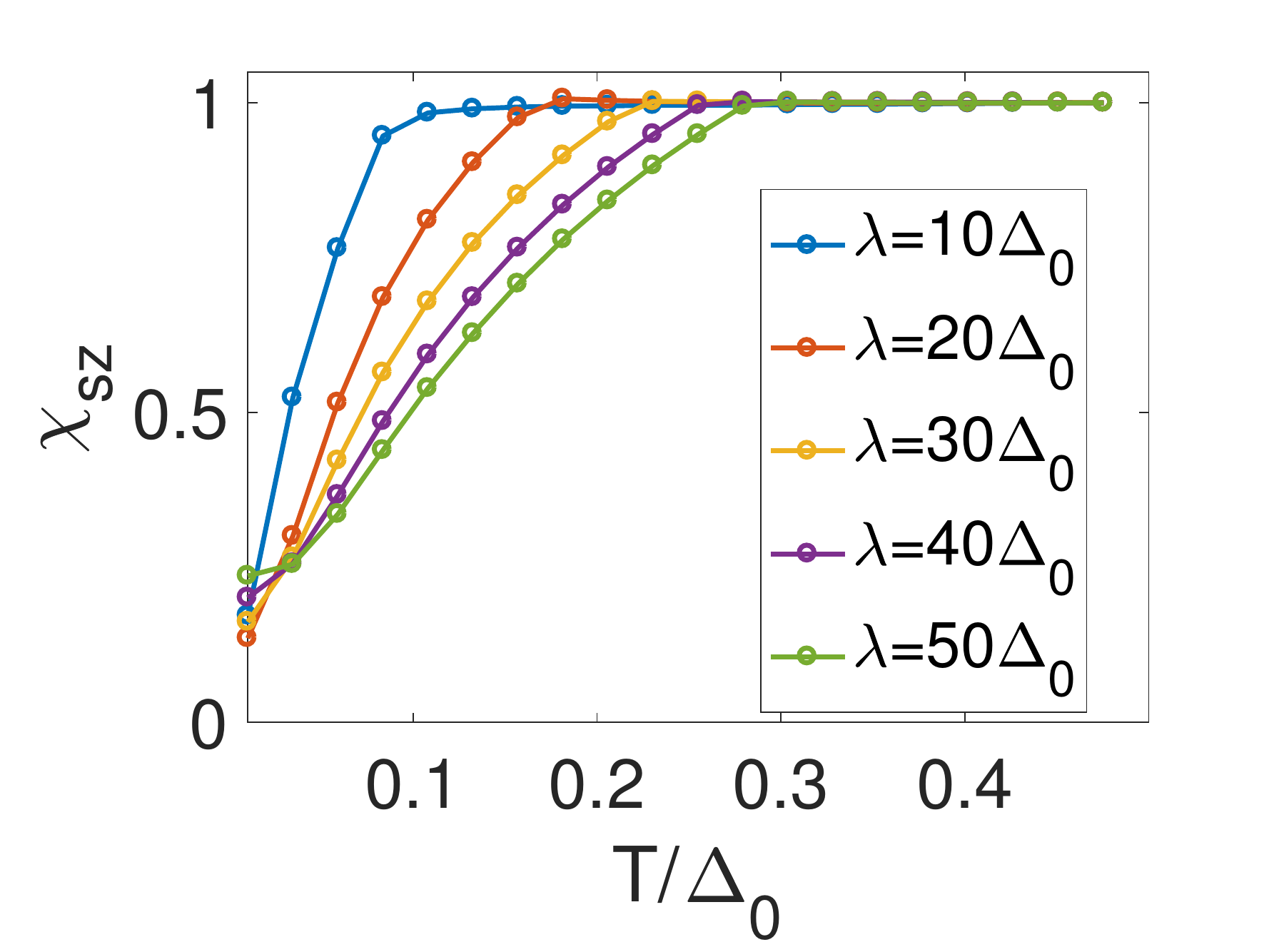}
\caption{Orbital susceptibility, spin susceptibility and gap function for triplet pairing $\Delta(\vec{k})=L_z\otimes{i\sigma_y\sigma_z}$ when adding $z$-direction magnetic field in the large SOC regime.}\label{F3}
\end{figure}

We now consider the regime with large SOC. The induced spin-singlet pairing state greatly reduces the spin-susceptibility and the Knight shift in every direction, as shown in Fig.~\ref{F3}. The induced intra-band pairing provides sufficient superconducting instability, leading to significant enhancement of the critical temperature, as predicted in the theoretical work~\cite{vafek1}. 

\begin{figure}[htb]
\centering
\includegraphics[width=4cm]{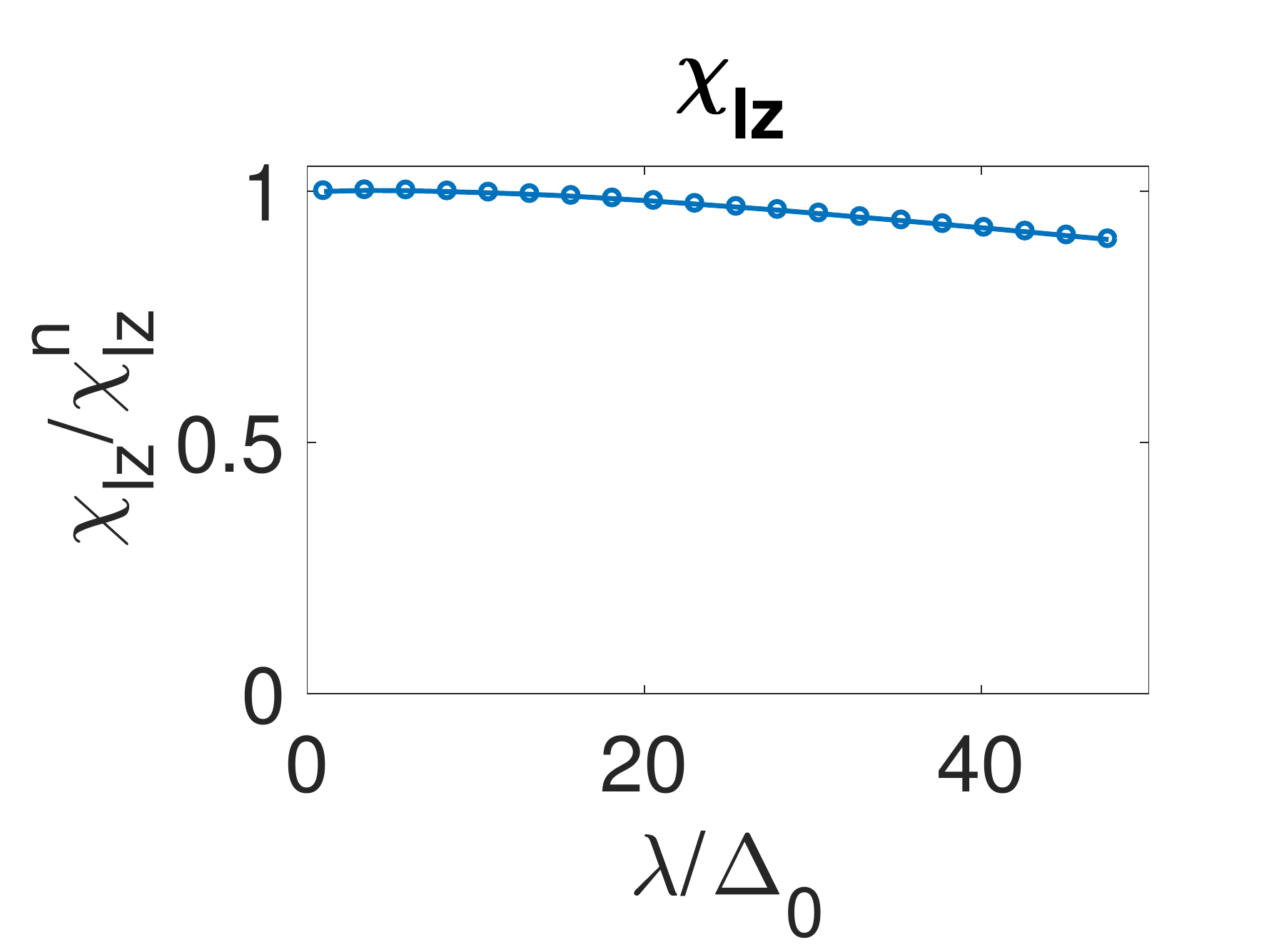}
\includegraphics[width=4cm]{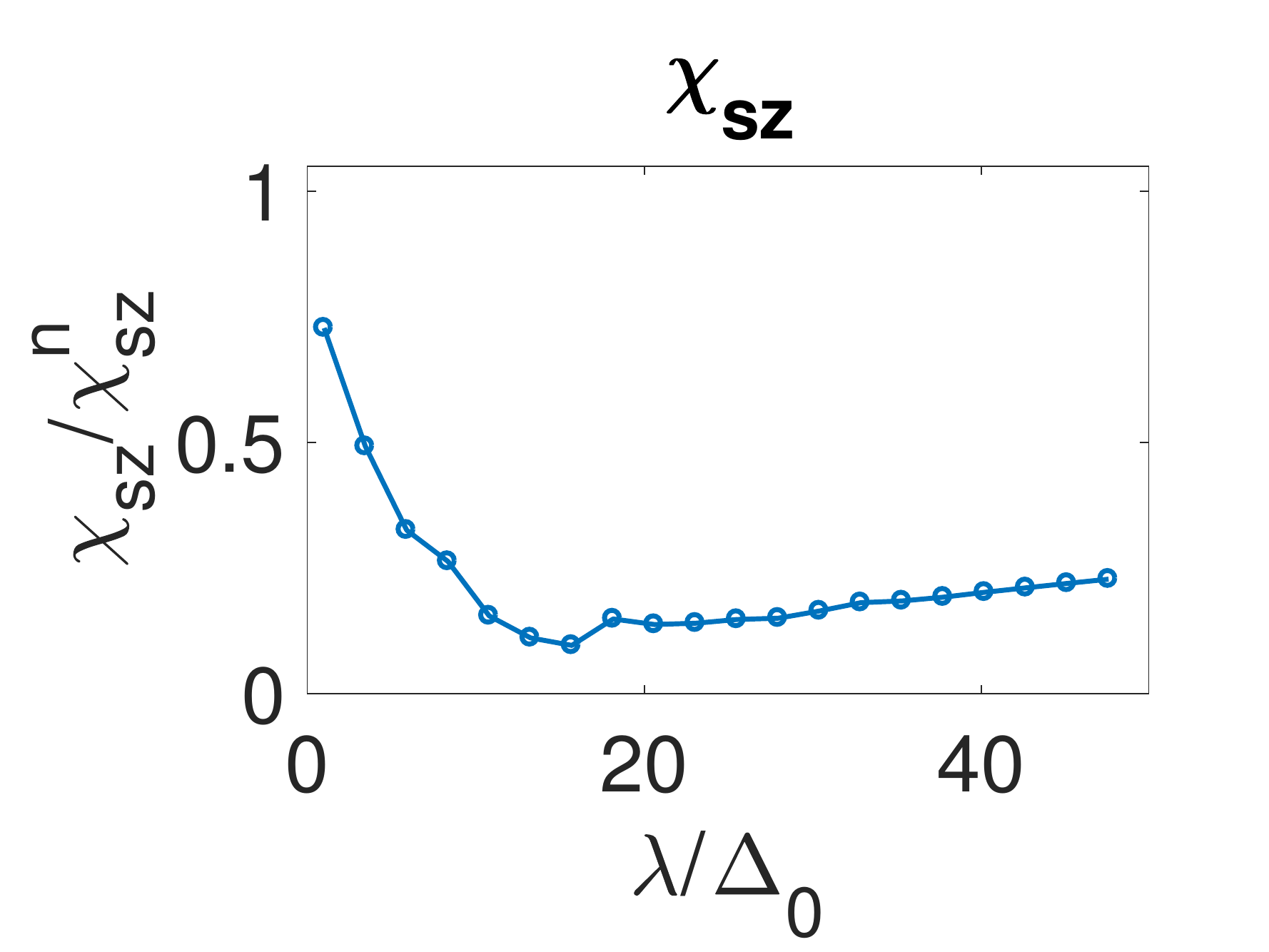}
\includegraphics[width=4cm]{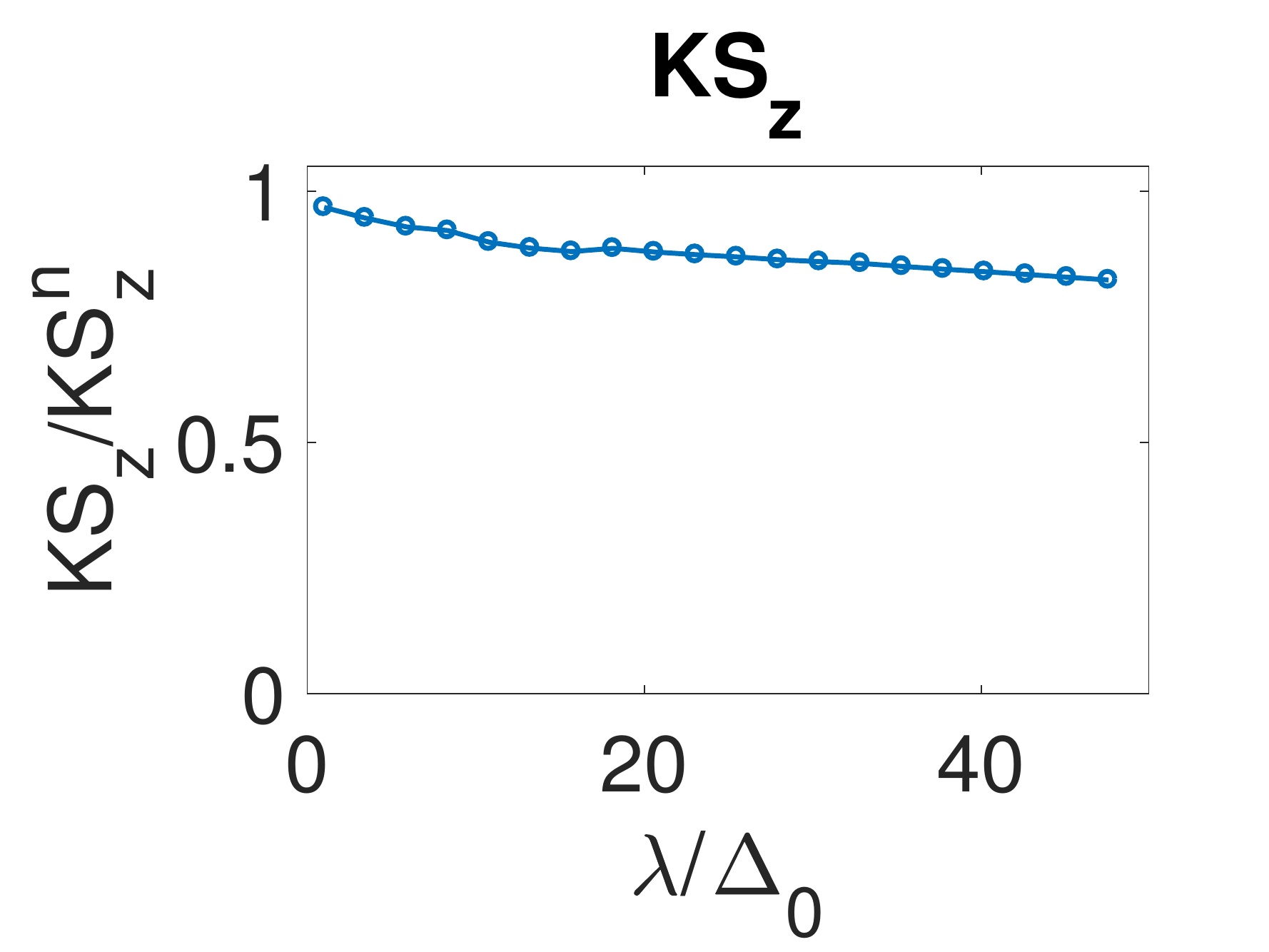}
\caption{Residual Knight shift and susceptibilities for an applied magnetic field in the $z$-direction, as a function of SOC. The strength of SOC varies from 0 to $50\Delta_{BCS}$, i.e. in a regime of strong SOC.}
\label{F4}
\end{figure}

\begin{figure}[htb]
\centering
\includegraphics[width=4cm]{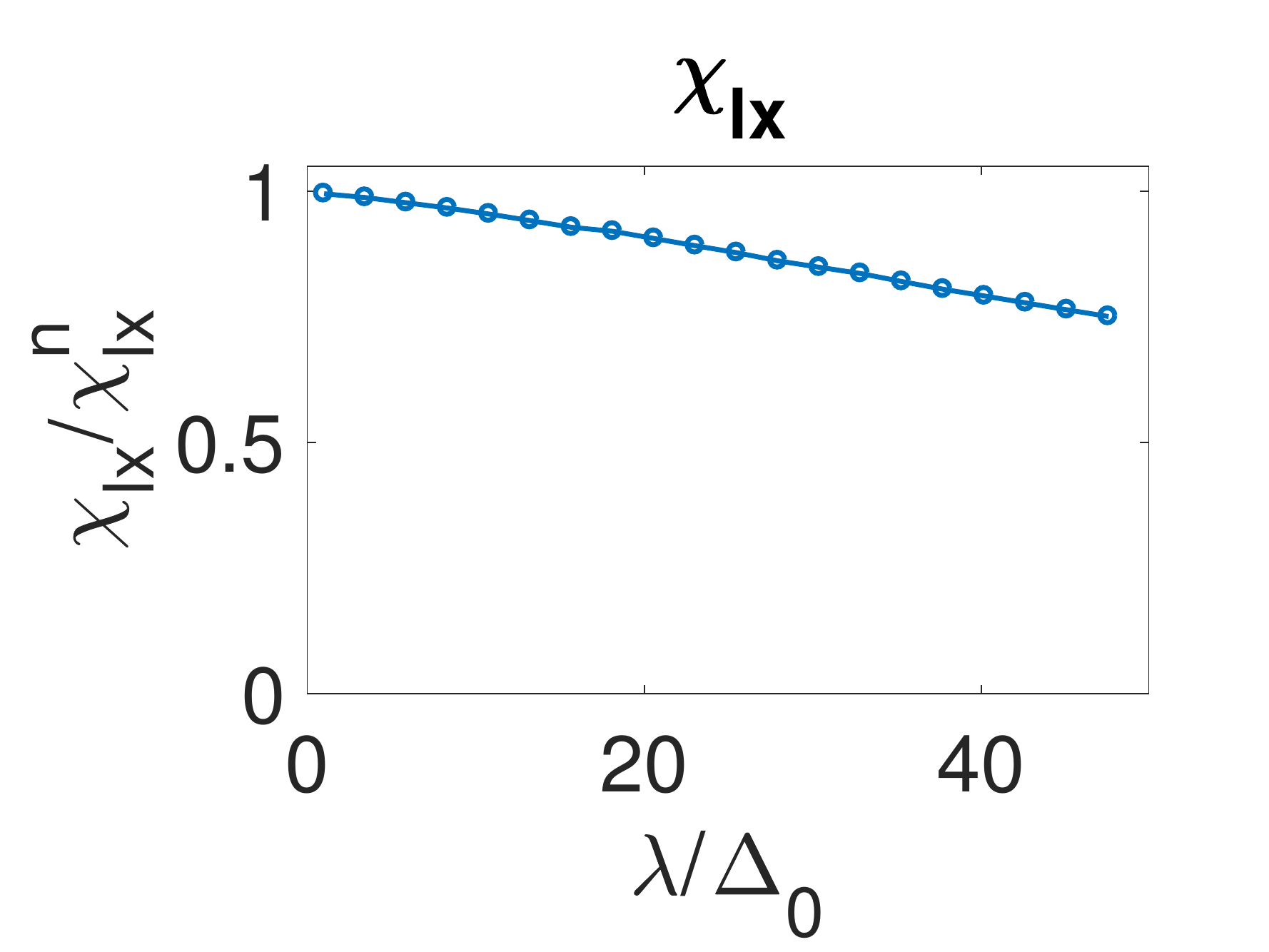}
\includegraphics[width=4cm]{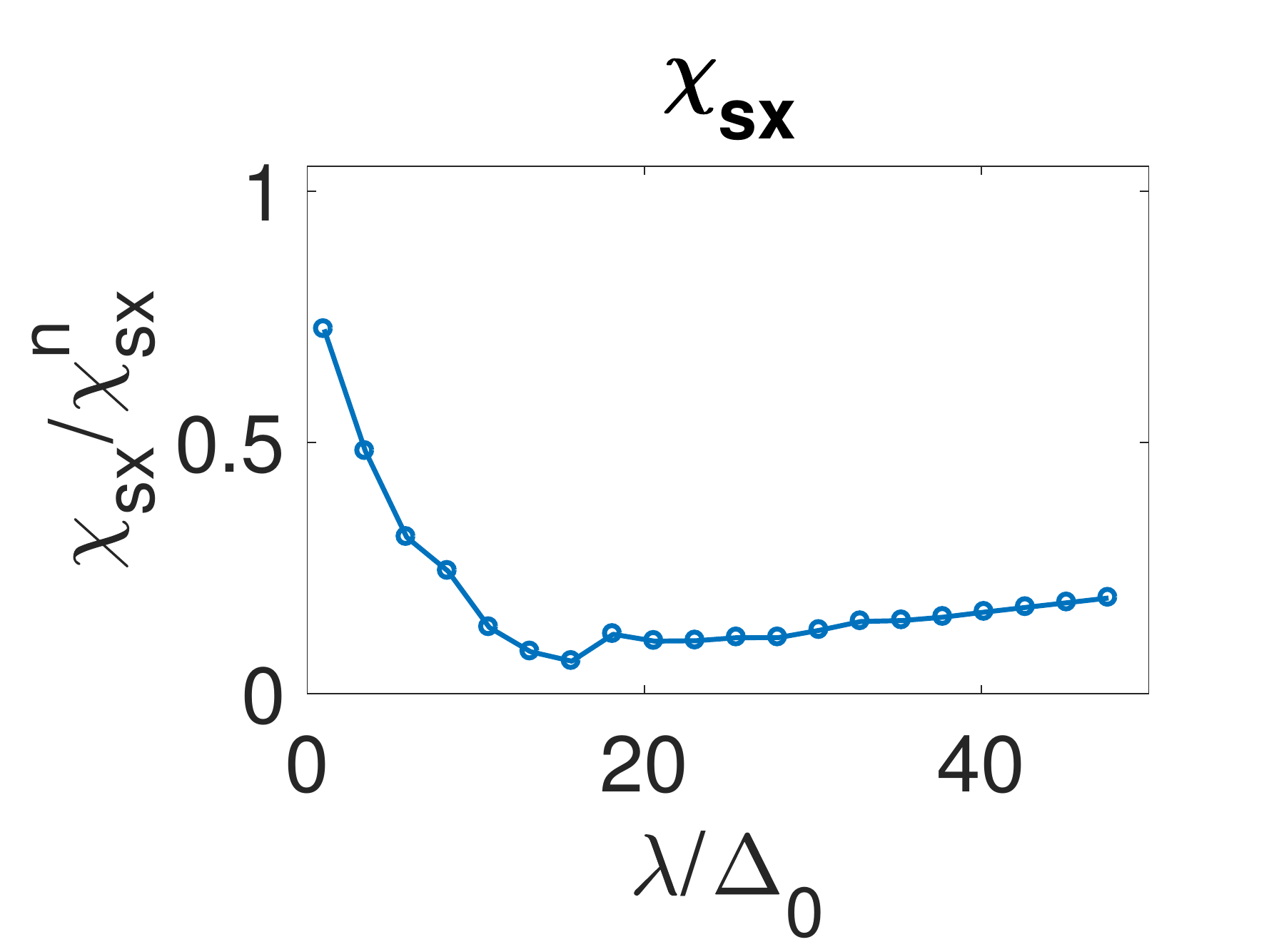}
\includegraphics[width=4cm]{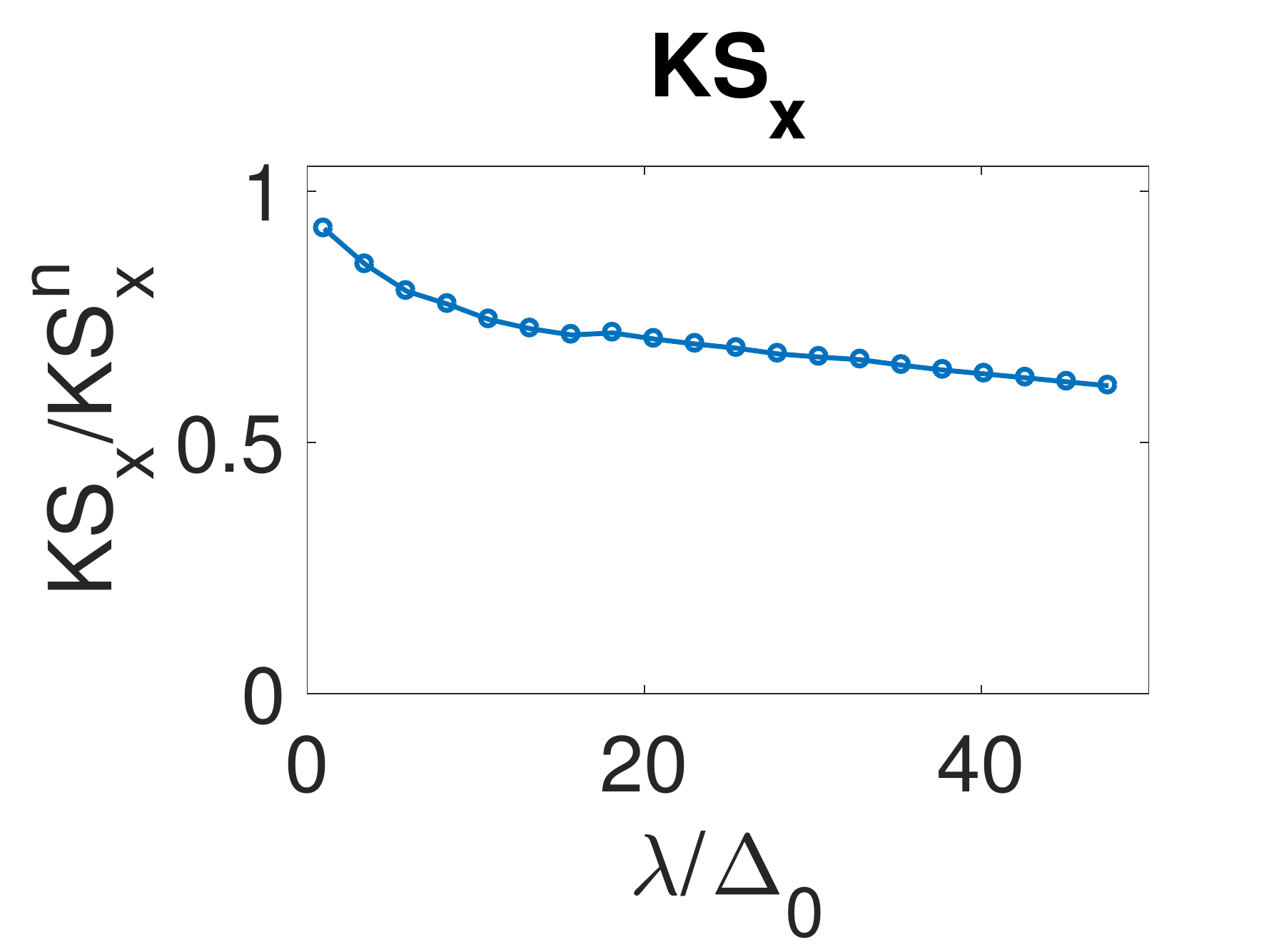}
\caption{Residual Knight shift and susceptibilities under a magnetic field applied in the $x$ direction, as a function of SOC strength. The strength of SOC varies from 0 to $50\Delta_{BCS}$.}\label{F5}
\end{figure}

In Fig.~\ref{F4} and Fig.~\ref{F5}, we focus on the residual susceptibilities and Knight shift for different SOC. We compare the results at two temperature.  At the lower temperature $T=0.01\Delta_{BCS}$, we add a BCS gap $\Delta_{BCS}=0.01$. At the ``higher'' temperature $T=\Delta_{BCS}$, we consider the normal state (with zero BCS gap). We take the ratio of the susceptibility in the first state to its value in the second state as a measure of the residual susceptibility. The aim is to observe the residual susceptibility at low temperature as we tune both the BCS term and SOC strength.

The residual spin susceptibility first exhibits a continuous drop as SOC is increased from zero, due to the formation of intra-band pairing. It reaches a minimum at around $\lambda=10\Delta_{BCS}$. More importantly, the residual spin susceptibility never goes to zero, even when the orbital susceptibility is approximately unchanged. This feature can be explained by returning to the pedagogical model with $\lambda=0$ (see Fig.~\ref{F3a} and Fig.~\ref{F3b} and accompanying discussion). This is a crucial difference compared with the well-studied $k_x+ik_y$ model, in which the non-zero spin susceptibility under SOC is accompanied by a decreasing orbital susceptibility \cite{PhysRevLett.92.097001,nisson2016nuclear}.

The non-zero spin susceptibility persists as the SOC is tuned to very large values. This may be due to mixing between an originally vanishing spin susceptibility and a non-zero orbital susceptibility, similar to the $k_x+ik_y$ model (in which the $z$-direction susceptibility exhibits similar features) or in a simple single-band spin-singlet model \cite{abrikosov1962spin,PhysRevLett.3.325}. The system becomes more complex due to band crossings. Therefore, for future work, the analysis for very large SOC could be performed for a model with a large band gap.

There is no qualitative difference between the drops in different directions, which confirms the predicted contribution from spin-singlet pairing when SOC is present \cite{vafek1}. Thus, the Knight shift in different directions provides a tool in identifying the inter-band state and the SOC strength. 

\section{conclusion} \label{sect:conc}
In summary, we have calculated the Knight shift for the even-parity orbital-singlet spin-triplet superconductors in a quasi-two dimensional tight binding model, by numerically solving the gap equation. In contrast to the $k_x+ik_y$ unconventional superconductor, the inter-band model exhibits no decrease in Knight shift in any direction. After introducing the required large spin orbit coupling, the predicted intra-band spin singlet state is observed, and there is a drop in spin susceptibility and Knight shift. However, the residual spin susceptibility is non-zero even under large SOC in contrast to spin-singlet intra-band pairing state. 

\section*{Acknowledgments}

Y.Y., A.C., and S.R. were supported by the DOE Office of Basic Energy Sciences, contract DEAC02- 76SF00515. D. F. A. was supported by the Gordon and Betty Moore Foundation's EPiQS Initiative through Grant No. GBMF4302.

\bibliography{citation}{}

\begin{thebibliography}{25}
\expandafter\ifx\csname natexlab\endcsname\relax\def\natexlab#1{#1}\fi
\expandafter\ifx\csname bibnamefont\endcsname\relax
  \def\bibnamefont#1{#1}\fi
\expandafter\ifx\csname bibfnamefont\endcsname\relax
  \def\bibfnamefont#1{#1}\fi
\expandafter\ifx\csname citenamefont\endcsname\relax
  \def\citenamefont#1{#1}\fi
\expandafter\ifx\csname url\endcsname\relax
  \def\url#1{\texttt{#1}}\fi
\expandafter\ifx\csname urlprefix\endcsname\relax\def\urlprefix{URL }\fi
\providecommand{\bibinfo}[2]{#2}
\providecommand{\eprint}[2][]{\url{#2}}

\bibitem[{\citenamefont{Scaffidi}(2017)}]{scaffidi2017thesis}
\bibinfo{author}{\bibfnamefont{T.}~\bibnamefont{Scaffidi}},
  \emph{\bibinfo{title}{Weak-Coupling Theory of Topological Superconductivity:
  The Case of Strontium Ruthenate}} (\bibinfo{year}{2017}), ISBN
  \bibinfo{isbn}{978-3-319-62866-0}.

\bibitem[{\citenamefont{Ishida et~al.}(1998)\citenamefont{Ishida, Mukuda,
  Kitaoka, Asayama, Mao, Mori, and Maeno}}]{ishida1998spin}
\bibinfo{author}{\bibfnamefont{K.}~\bibnamefont{Ishida}},
  \bibinfo{author}{\bibfnamefont{H.}~\bibnamefont{Mukuda}},
  \bibinfo{author}{\bibfnamefont{Y.}~\bibnamefont{Kitaoka}},
  \bibinfo{author}{\bibfnamefont{K.}~\bibnamefont{Asayama}},
  \bibinfo{author}{\bibfnamefont{Z.}~\bibnamefont{Mao}},
  \bibinfo{author}{\bibfnamefont{Y.}~\bibnamefont{Mori}}, \bibnamefont{and}
  \bibinfo{author}{\bibfnamefont{Y.}~\bibnamefont{Maeno}},
  \bibinfo{journal}{Nature} \textbf{\bibinfo{volume}{396}},
  \bibinfo{pages}{658} (\bibinfo{year}{1998}).

\bibitem[{\citenamefont{Xia et~al.}(2006)\citenamefont{Xia, Maeno, Beyersdorf,
  Fejer, and Kapitulnik}}]{xia1}
\bibinfo{author}{\bibfnamefont{J.}~\bibnamefont{Xia}},
  \bibinfo{author}{\bibfnamefont{Y.}~\bibnamefont{Maeno}},
  \bibinfo{author}{\bibfnamefont{P.~T.} \bibnamefont{Beyersdorf}},
  \bibinfo{author}{\bibfnamefont{M.~M.} \bibnamefont{Fejer}}, \bibnamefont{and}
  \bibinfo{author}{\bibfnamefont{A.}~\bibnamefont{Kapitulnik}},
  \bibinfo{journal}{Phys. Rev. Lett.} \textbf{\bibinfo{volume}{97}},
  \bibinfo{pages}{167002} (\bibinfo{year}{2006}),
  \urlprefix\url{https://link.aps.org/doi/10.1103/PhysRevLett.97.167002}.

\bibitem[{\citenamefont{Hicks et~al.}(2010)\citenamefont{Hicks, Kirtley,
  Lippman, Koshnick, Huber, Maeno, Yuhasz, Maple, and Moler}}]{hicks1}
\bibinfo{author}{\bibfnamefont{C.~W.} \bibnamefont{Hicks}},
  \bibinfo{author}{\bibfnamefont{J.~R.} \bibnamefont{Kirtley}},
  \bibinfo{author}{\bibfnamefont{T.~M.} \bibnamefont{Lippman}},
  \bibinfo{author}{\bibfnamefont{N.~C.} \bibnamefont{Koshnick}},
  \bibinfo{author}{\bibfnamefont{M.~E.} \bibnamefont{Huber}},
  \bibinfo{author}{\bibfnamefont{Y.}~\bibnamefont{Maeno}},
  \bibinfo{author}{\bibfnamefont{W.~M.} \bibnamefont{Yuhasz}},
  \bibinfo{author}{\bibfnamefont{M.~B.} \bibnamefont{Maple}}, \bibnamefont{and}
  \bibinfo{author}{\bibfnamefont{K.~A.} \bibnamefont{Moler}},
  \bibinfo{journal}{Phys. Rev. B} \textbf{\bibinfo{volume}{81}},
  \bibinfo{pages}{214501} (\bibinfo{year}{2010}),
  \urlprefix\url{https://link.aps.org/doi/10.1103/PhysRevB.81.214501}.

\bibitem[{\citenamefont{Steppke et~al.}(2017)\citenamefont{Steppke, Zhao,
  Barber, Scaffidi, Jerzembeck, Rosner, Gibbs, Maeno, Simon, Mackenzie
  et~al.}}]{steppke2017}
\bibinfo{author}{\bibfnamefont{A.}~\bibnamefont{Steppke}},
  \bibinfo{author}{\bibfnamefont{L.}~\bibnamefont{Zhao}},
  \bibinfo{author}{\bibfnamefont{M.~E.} \bibnamefont{Barber}},
  \bibinfo{author}{\bibfnamefont{T.}~\bibnamefont{Scaffidi}},
  \bibinfo{author}{\bibfnamefont{F.}~\bibnamefont{Jerzembeck}},
  \bibinfo{author}{\bibfnamefont{H.}~\bibnamefont{Rosner}},
  \bibinfo{author}{\bibfnamefont{A.~S.} \bibnamefont{Gibbs}},
  \bibinfo{author}{\bibfnamefont{Y.}~\bibnamefont{Maeno}},
  \bibinfo{author}{\bibfnamefont{S.~H.} \bibnamefont{Simon}},
  \bibinfo{author}{\bibfnamefont{A.~P.} \bibnamefont{Mackenzie}},
  \bibnamefont{et~al.}, \bibinfo{journal}{Science}
  \textbf{\bibinfo{volume}{355}}, \bibinfo{pages}{eaaf9398}
  (\bibinfo{year}{2017}).

\bibitem[{\citenamefont{Vafek and Chubukov}(2017)}]{vafek1}
\bibinfo{author}{\bibfnamefont{O.}~\bibnamefont{Vafek}} \bibnamefont{and}
  \bibinfo{author}{\bibfnamefont{A.~V.} \bibnamefont{Chubukov}},
  \bibinfo{journal}{Phys. Rev. Lett.} \textbf{\bibinfo{volume}{118}},
  \bibinfo{pages}{087003} (\bibinfo{year}{2017}),
  \urlprefix\url{https://link.aps.org/doi/10.1103/PhysRevLett.118.087003}.

\bibitem[{\citenamefont{Ramires and Sigrist}(2016)}]{ramires2}
\bibinfo{author}{\bibfnamefont{A.}~\bibnamefont{Ramires}} \bibnamefont{and}
  \bibinfo{author}{\bibfnamefont{M.}~\bibnamefont{Sigrist}},
  \bibinfo{journal}{Phys. Rev. B} \textbf{\bibinfo{volume}{94}},
  \bibinfo{pages}{104501} (\bibinfo{year}{2016}),
  \urlprefix\url{https://link.aps.org/doi/10.1103/PhysRevB.94.104501}.

\bibitem[{\citenamefont{Hoshino and Werner}(2015)}]{hoshino1}
\bibinfo{author}{\bibfnamefont{S.}~\bibnamefont{Hoshino}} \bibnamefont{and}
  \bibinfo{author}{\bibfnamefont{P.}~\bibnamefont{Werner}},
  \bibinfo{journal}{Phys. Rev. Lett.} \textbf{\bibinfo{volume}{115}},
  \bibinfo{pages}{247001} (\bibinfo{year}{2015}),
  \urlprefix\url{https://link.aps.org/doi/10.1103/PhysRevLett.115.247001}.

\bibitem[{\citenamefont{Borisenko et~al.}(2016)\citenamefont{Borisenko,
  Evtushinsky, Liu, Morozov, Kappenberger, Wurmehl, B{\"u}chner, Yaresko, Kim,
  Hoesch et~al.}}]{borisenko2016direct}
\bibinfo{author}{\bibfnamefont{S.~V.} \bibnamefont{Borisenko}},
  \bibinfo{author}{\bibfnamefont{D.}~\bibnamefont{Evtushinsky}},
  \bibinfo{author}{\bibfnamefont{Z.-H.} \bibnamefont{Liu}},
  \bibinfo{author}{\bibfnamefont{I.}~\bibnamefont{Morozov}},
  \bibinfo{author}{\bibfnamefont{R.}~\bibnamefont{Kappenberger}},
  \bibinfo{author}{\bibfnamefont{S.}~\bibnamefont{Wurmehl}},
  \bibinfo{author}{\bibfnamefont{B.}~\bibnamefont{B{\"u}chner}},
  \bibinfo{author}{\bibfnamefont{A.}~\bibnamefont{Yaresko}},
  \bibinfo{author}{\bibfnamefont{T.}~\bibnamefont{Kim}},
  \bibinfo{author}{\bibfnamefont{M.}~\bibnamefont{Hoesch}},
  \bibnamefont{et~al.}, \bibinfo{journal}{Nature Physics}
  \textbf{\bibinfo{volume}{12}}, \bibinfo{pages}{311} (\bibinfo{year}{2016}).

\bibitem[{\citenamefont{Nisson and Curro}(2016)}]{nisson2016nuclear}
\bibinfo{author}{\bibfnamefont{D.}~\bibnamefont{Nisson}} \bibnamefont{and}
  \bibinfo{author}{\bibfnamefont{N.}~\bibnamefont{Curro}},
  \bibinfo{journal}{New Journal of Physics} \textbf{\bibinfo{volume}{18}},
  \bibinfo{pages}{073041} (\bibinfo{year}{2016}).

\bibitem[{\citenamefont{Haverkort et~al.}(2008)\citenamefont{Haverkort,
  Elfimov, Tjeng, Sawatzky, and Damascelli}}]{haverkort1}
\bibinfo{author}{\bibfnamefont{M.~W.} \bibnamefont{Haverkort}},
  \bibinfo{author}{\bibfnamefont{I.~S.} \bibnamefont{Elfimov}},
  \bibinfo{author}{\bibfnamefont{L.~H.} \bibnamefont{Tjeng}},
  \bibinfo{author}{\bibfnamefont{G.~A.} \bibnamefont{Sawatzky}},
  \bibnamefont{and}
  \bibinfo{author}{\bibfnamefont{A.}~\bibnamefont{Damascelli}},
  \bibinfo{journal}{Phys. Rev. Lett.} \textbf{\bibinfo{volume}{101}},
  \bibinfo{pages}{026406} (\bibinfo{year}{2008}),
  \urlprefix\url{https://link.aps.org/doi/10.1103/PhysRevLett.101.026406}.

\bibitem[{\citenamefont{Ramires et~al.}(2018)\citenamefont{Ramires, Agterberg,
  and Sigrist}}]{ramires1}
\bibinfo{author}{\bibfnamefont{A.}~\bibnamefont{Ramires}},
  \bibinfo{author}{\bibfnamefont{D.~F.} \bibnamefont{Agterberg}},
  \bibnamefont{and} \bibinfo{author}{\bibfnamefont{M.}~\bibnamefont{Sigrist}},
  \bibinfo{journal}{arXiv preprint arXiv:1802.00361}  (\bibinfo{year}{2018}).

\bibitem[{\citenamefont{Klejnberg and Spalek}(1999)}]{klejnberg1999}
\bibinfo{author}{\bibfnamefont{A.}~\bibnamefont{Klejnberg}} \bibnamefont{and}
  \bibinfo{author}{\bibfnamefont{J.}~\bibnamefont{Spalek}},
  \bibinfo{journal}{Journal of Physics: Condensed Matter}
  \textbf{\bibinfo{volume}{11}}, \bibinfo{pages}{6553} (\bibinfo{year}{1999}).

\bibitem[{\citenamefont{Spa\l{}ek}(2001)}]{spalek2001}
\bibinfo{author}{\bibfnamefont{J.}~\bibnamefont{Spa\l{}ek}},
  \bibinfo{journal}{Phys. Rev. B} \textbf{\bibinfo{volume}{63}},
  \bibinfo{pages}{104513} (\bibinfo{year}{2001}),
  \urlprefix\url{https://link.aps.org/doi/10.1103/PhysRevB.63.104513}.

\bibitem[{\citenamefont{Han}(2004)}]{han2004}
\bibinfo{author}{\bibfnamefont{J.~E.} \bibnamefont{Han}},
  \bibinfo{journal}{Phys. Rev. B} \textbf{\bibinfo{volume}{70}},
  \bibinfo{pages}{054513} (\bibinfo{year}{2004}),
  \urlprefix\url{https://link.aps.org/doi/10.1103/PhysRevB.70.054513}.

\bibitem[{\citenamefont{Mackenzie and
  Maeno}(2003)}]{mackenzie2003superconductivity}
\bibinfo{author}{\bibfnamefont{A.~P.} \bibnamefont{Mackenzie}}
  \bibnamefont{and} \bibinfo{author}{\bibfnamefont{Y.}~\bibnamefont{Maeno}},
  \bibinfo{journal}{Reviews of Modern Physics} \textbf{\bibinfo{volume}{75}},
  \bibinfo{pages}{657} (\bibinfo{year}{2003}).

\bibitem[{\citenamefont{Abragam}(1961)}]{abragam1961principles}
\bibinfo{author}{\bibfnamefont{A.}~\bibnamefont{Abragam}},
  \emph{\bibinfo{title}{The principles of nuclear magnetism}},
  \bibinfo{number}{32} (\bibinfo{publisher}{Oxford university press},
  \bibinfo{year}{1961}).

\bibitem[{\citenamefont{Curro}(2009)}]{curro2009nuclear}
\bibinfo{author}{\bibfnamefont{N.}~\bibnamefont{Curro}},
  \bibinfo{journal}{Reports on Progress in Physics}
  \textbf{\bibinfo{volume}{72}}, \bibinfo{pages}{026502}
  (\bibinfo{year}{2009}).

\bibitem[{\citenamefont{Rigamonti et~al.}(1998)\citenamefont{Rigamonti, Borsa,
  and Carretta}}]{rigamonti1998basic}
\bibinfo{author}{\bibfnamefont{A.}~\bibnamefont{Rigamonti}},
  \bibinfo{author}{\bibfnamefont{F.}~\bibnamefont{Borsa}}, \bibnamefont{and}
  \bibinfo{author}{\bibfnamefont{P.}~\bibnamefont{Carretta}},
  \bibinfo{journal}{Reports on Progress in Physics}
  \textbf{\bibinfo{volume}{61}}, \bibinfo{pages}{1367} (\bibinfo{year}{1998}).

\bibitem[{\citenamefont{Frigeri et~al.}(2004)\citenamefont{Frigeri, Agterberg,
  Koga, and Sigrist}}]{PhysRevLett.92.097001}
\bibinfo{author}{\bibfnamefont{P.~A.} \bibnamefont{Frigeri}},
  \bibinfo{author}{\bibfnamefont{D.~F.} \bibnamefont{Agterberg}},
  \bibinfo{author}{\bibfnamefont{A.}~\bibnamefont{Koga}}, \bibnamefont{and}
  \bibinfo{author}{\bibfnamefont{M.}~\bibnamefont{Sigrist}},
  \bibinfo{journal}{Phys. Rev. Lett.} \textbf{\bibinfo{volume}{92}},
  \bibinfo{pages}{097001} (\bibinfo{year}{2004}),
  \urlprefix\url{https://link.aps.org/doi/10.1103/PhysRevLett.92.097001}.

\bibitem[{\citenamefont{Akbari and Thalmeier}(2013)}]{band}
\bibinfo{author}{\bibfnamefont{A.}~\bibnamefont{Akbari}} \bibnamefont{and}
  \bibinfo{author}{\bibfnamefont{P.}~\bibnamefont{Thalmeier}},
  \bibinfo{journal}{Phys. Rev. B} \textbf{\bibinfo{volume}{88}},
  \bibinfo{pages}{134519} (\bibinfo{year}{2013}),
  \urlprefix\url{https://link.aps.org/doi/10.1103/PhysRevB.88.134519}.

\bibitem[{\citenamefont{Abrikosov and Gor’kov}(1962)}]{abrikosov1962spin}
\bibinfo{author}{\bibfnamefont{A.}~\bibnamefont{Abrikosov}} \bibnamefont{and}
  \bibinfo{author}{\bibfnamefont{L.}~\bibnamefont{Gor’kov}},
  \bibinfo{journal}{Sov. Phys. JETP} \textbf{\bibinfo{volume}{15}},
  \bibinfo{pages}{752} (\bibinfo{year}{1962}).

\bibitem[{\citenamefont{Anderson}(1959)}]{PhysRevLett.3.325}
\bibinfo{author}{\bibfnamefont{P.~W.} \bibnamefont{Anderson}},
  \bibinfo{journal}{Phys. Rev. Lett.} \textbf{\bibinfo{volume}{3}},
  \bibinfo{pages}{325} (\bibinfo{year}{1959}),
  \urlprefix\url{https://link.aps.org/doi/10.1103/PhysRevLett.3.325}.

\bibitem[{\citenamefont{Abragam and Bleaney}(2012)}]{NMR}
\bibinfo{author}{\bibfnamefont{A.}~\bibnamefont{Abragam}} \bibnamefont{and}
  \bibinfo{author}{\bibfnamefont{B.}~\bibnamefont{Bleaney}},
  \emph{\bibinfo{title}{Electron paramagnetic resonance of transition ions}}
  (\bibinfo{publisher}{OUP Oxford}, \bibinfo{year}{2012}).

\bibitem[{\citenamefont{Griffith}(1971)}]{chivalue}
\bibinfo{author}{\bibfnamefont{J.~S.} \bibnamefont{Griffith}},
  \emph{\bibinfo{title}{The theory of transition-metal ions}}
  (\bibinfo{publisher}{Cambridge University Press}, \bibinfo{year}{1971}).

\end{thebibliography}

\end{document}